\documentclass[11pt, authoryear]{elsarticle}              % for format 
\journal{Transportation Research Part A: Policy and Practice}            % 
\usepackage{amssymb,amsfonts,amsmath}
\usepackage{xcolor}

\usepackage{natbib}
\usepackage[top=1.0in, bottom=1.0in, left=1.0in, right=1.0 in]{geometry}   %
\usepackage[onehalfspacing]{setspace}
\usepackage[bottom]{footmisc}
\usepackage{indentfirst}
\usepackage{endnotes}
\usepackage{verbatim}                % for block comments
\usepackage{lastpage,placeins}
\usepackage{hyperref}                % for emails. Also, it links all your
\usepackage{graphicx,epsfig,epstopdf,subfig}       % for including eps,jpeg
\usepackage{array}
\usepackage{afterpage,rotating,longtable,booktabs, colortbl}       % needed 
\usepackage{mdwlist}
\usepackage{pdflscape}
\usepackage{enumerate}
\usepackage[shortlabels]{enumitem}
\usepackage{soul}
\usepackage{appendix}
\usepackage{breakcites}
\usepackage{bm}

\usepackage[standard]{ntheorem}
\usepackage[algo2e]{algorithm2e}

\allowdisplaybreaks[1]                     % page breaks over multi-line 
\theoremstyle{break}
\theorembodyfont{\normalfont}

\newcommand*{\xdash}[1][3em]{\rule[0.5ex]{#1}{1.25pt}}
\newcommand*{\bxdash}[1][1em]{\rule[0.5ex]{#1}{1.25pt}}

\begin{document}

\begin{frontmatter}
\title{\textbf{Do Determinants of EV Purchase Intent vary across the
Spectrum? Evidence from Bayesian Analysis of US Survey Data} 
}

\author{Nafisa Lohawala}
\ead{nlohawala@rff.org}
\address{Resources for the Future, Washington DC, USA}
 
\author{Mohammad Arshad Rahman\corref{cor1}}
\ead{marshad@iitk.ac.in}
%\fntext[fn1]{Email address: marshad@iitk.ac.in. }
\address{Department of Economic Sciences, Indian Institute of Technology 
Kanpur, India.} % \\
%              Office: Faculty Building 672, IITK, Kanpur.
%              Phone: +91 969-564-7700. Fax: +91 512-259-7570.}
\cortext[cor1]{Corresponding author}

\begin{abstract}
While electric vehicle (EV) adoption has been widely studied, most research 
focuses on the average effects of predictors on purchase intent, overlooking 
variation across the distribution of EV purchase intent. This paper makes a 
threefold contribution by analyzing four unique explanatory variables, 
leveraging large-scale US survey data from 2021 to 2023, and employing 
Bayesian ordinal probit and Bayesian ordinal quantile modeling to evaluate 
the effects of these variables$-$while controlling for other commonly used 
covariates$-$on EV purchase intent, both on average and across its full 
distribution. By modeling purchase intent as an ordered 
outcome$-$from “not at all likely” to “very likely”$-$we reveal how covariate 
effects differ across levels of interest. This is the first application of 
ordinal quantile modeling in the EV adoption literature, uncovering 
heterogeneity in how potential buyers respond to key factors. For instance, 
confidence in development of charging infrastructure and belief in 
environmental benefits are linked not only to higher interest among likely 
adopters but also to reduced resistance among more skeptical respondents. 
Notably, we identify a gap between the prevalence and influence of key 
predictors: although few respondents report strong infrastructure confidence 
or frequent EV information exposure, both factors are strongly 
associated with increased intent across the spectrum. These findings suggest 
clear opportunities for targeted communication and outreach, alongside 
infrastructure investment, to support widespread EV adoption.
\end{abstract}

\end{frontmatter}

%------------------------------------------------------------------------------
\section{Introduction}\label{sec:Intro}
%------------------------------------------------------------------------------

Transportation is the largest source of greenhouse gas emissions in the 
United States (US), representing 28\% of the total emissions in 2022, with 
light-duty vehicles responsible for more than half \citep{EPA-2024}. Electric 
vehicles (EVs),  given their zero tailpipe emissions, are widely regarded as 
a key solution for reducing emissions and meeting climate targets 
\citep{Bhat-etal-2025}. Over the past decade, governments worldwide, 
including the US, have implemented a range of policies and incentives aimed 
at encouraging EV adoption \citep{jenn2018effectiveness, Mekky-Collins-2024}. 
Key US measures have included nationwide greenhouse gas standards to reduce 
emissions from new vehicles; up to \$7,500 federal tax credits for EV buyers; 
investments in charging infrastructure; and state-level mandates like 
California's Zero-Emission Vehicle (ZEV) program, which requires automakers 
to sell an increasing share of EVs\footnote{For more details, see the EPA's 
\href{https://www.epa.gov/regulations-emissions-vehicles-and-engines/light-duty-vehicle-greenhouse-gas-regulations-and}
{greenhouse gas regulations}, the IRS 
\href{https://www.irs.gov/credits-deductions/credits-for-new-clean-vehicles-purchased-in-2023-or-after}
{EV tax credit guidelines}, National Electric Vehicle Infrastructure   
(\href{https://www.fhwa.dot.gov/environment/nevi/https://www.fhwa.dot.gov/environment/nevi/}{NEVI})
 Program and California's 
\href{https://ww2.arb.ca.gov/our-work/programs/zero-emission-vehicle-program}{ZEV
 Program}. As of early 2026, several of these EV-focused policies have been 
 rolled back under the Trump administration.}.

Despite these efforts, US sales remain relatively low, accounting 
for only 10\% of global new EV sales in 2023, trailing behind China 
(60\%) and Europe (25\%) \citep{IEA-2024}. The 
trajectory of US adoption is also increasingly uncertain. In 
2024, several automakers, including GM and Ford, scaled back ambitious EV 
targets, citing weak consumer demand \citep{Colias-Otts-2024}. Successive administrations have taken sharply different approaches 
to EVs, creating uncertainty around long-term incentives and regulations. The 
Biden administration promoted EV adoption through subsidies, 
infrastructure investments, and emissions regulations as part of its broader 
climate agenda. In contrast, the Trump administration has pursued a 
deregulatory shift: as of early 2026, federal EV tax credits have been 
eliminated, the EPA's endangerment finding 
is under reconsideration, and California's Clean Air Act waiver for 
state-level ZEV mandates has been revoked.\footnote{For details on the EV tax credit phase-out, see Section 70502 of the \href{https://www.congress.gov/bill/119th-congress/house-bill/1/text}{One Big Beautiful Bill Act}. For details on the status of California's waiver for ZEV mandates, see \href{https://www.congress.gov/bill/119th-congress/house-joint-resolution/88/text}{H.J. Res. 88}. For details on the reconsideration of the endangerment finding, see the \href{https://www.epa.gov/regulations-emissions-vehicles-and-engines/proposed-rule-reconsideration-2009-endangerment-finding}{EPA website}. Federal charging-infrastructure funding was also suspended but is slated for restoration following litigation (see \href{https://www.transportation.gov/briefing-room/president-trumps-transportation-secretary-sean-p-duffy-unveils-revised-nevi-guidance}{Department of Transportation newsroom}). Other policy changes are subject to ongoing 
or possible future legal challenges, so further changes remain possible. 
}
These changing political priorities reflect tensions between 
environmental regulations, industry interests, and fiscal considerations in 
promoting EV adoption. Although EVs reduce emissions---particularly in 
regions with cleaner electricity grids---subsidies and incentives, along with 
declining gas tax revenue, pose budgetary challenges \citep{Banta-etal-2024}. 
The Trump administration has also cited concerns that stringent emissions and 
fuel economy regulations increase compliance costs for automakers and limit 
consumer choice. 

Amid fluctuating incentives and persistent regulatory uncertainty, EV 
adoption ultimately depends on consumer interest. Accordingly, the study of 
public opinion on EV adoption is particularly important, as public opinion is 
known to influence US public policy even in the presence of competing 
interests and partisan conflict \citep{Burstein-2003, Shapiro-2011}. While 
the literature on EV adoption is extensive, important knowledge gaps remain. 
We organize these gaps, and our contributions to addressing them, along three 
dimensions: (a) the covariates (independent variables 
or regressors) considered, (b) the scale and representativeness of the data, 
and (c) the statistical models used to examine the relationship between EV 
adoption and these covariates.

Within the first dimension, EV adoption is typically modeled 
as a function of demographic characteristics--such as age, gender, education, 
race, political affiliation, and income--together with factors 
related to cost \citep{liu2021comparing, woody2024electric}, convenience and 
charging infrastructure \citep{Chakraborty-etal-2019, Chakraborty-etal-2022}, 
perception of environmental benefits \citep{holland2016there, Lakshman-2024}, 
and awareness of these trade-offs \citep{Rezvani-etal-2015}. In this context, 
any determinant that plausibly affects EV adoption but has not been 
systematically examined represents an opportunity to deepen our understanding 
of adoption behavior. Building on this literature, we examine EV adoption 
using standard demographic controls and introduce four additional factors: 
confidence in building charging stations and infrastructure, perception of 
government climate efforts, exposure to EV-related information, and 
perceptions of EVs' environmental benefits relative to gasoline powered 
vehicles.

The first factor--confidence in development of future charging 
infrastructure--is crucial for EV adoption because purchase intent is likely 
to depend on expectations about future network expansion, particularly when 
the current system remains incomplete. This stands in contrast to existing 
studies such as \citet{Carley-etal-2013} and \citet{tiwari2020public}, which
emphasize the role of existing infrastructure. The second factor concerns
perceptions of government climate efforts. While previous research has 
documented links between EV purchase intent and political affiliation (e.g., 
\citet{mamkhezri2025public, sintov2020partisan}), it has not paid attention 
to individuals' assessments of government climate action itself. Because EVs 
are widely recognized as a climate policy 
instrument, individuals who believe the government is already doing enough 
may be less inclined to purchase an EV---particularly if they are less aware 
of EVs' ancillary benefits.

The third factor is exposure to EV-related information. Previous research has 
examined the effects of information on total cost of ownership 
\citep{dumortier2015effects}, media coverage \citep{Scherrer-2023}, and peer 
effects \citep{zhao2022determines}. We extend these works by using a 
nationally representative survey that directly asks respondents how much they 
have heard or read about EVs; although subjective, this measure reduces 
problems of indirect inference. The fourth factor is perceptions of EVs' 
environmental benefits. Several studies have considered consumers' 
pro-environmental attitudes \citep{hu2023policy, Carley-etal-2013, 
Krause-etal-2013, Khatua-etal-2023, Scherrer-2023, mamkhezri2025public}  and 
reputational concerns \citep{Buhmann-Criado-2023}. However, research on 
perceptions of EV's environmental benefits relative to gasoline powered 
vehicles is limited. \citet{lashari2021consumers} and \citet{Jia-etal-2025} 
examine this factor in a small sample, whereas our analysis situates it in a 
broader US context.\footnote{Studies have also examined other factors such as 
range anxiety, performance concerns, and consumer incentives (See 
\citet{Carley-etal-2013, tiwari2020public, zhao2022determines, 
stekelberg2024effect, mamkhezri2025public} among others). A separate strand 
of research examines determinants of actual EV sales, including tax 
credits and rebates \citep{deshazo2017designing, jenn2018effectiveness,
clinton2019providing, sheldon2023electric}, charging infrastructure 
\citep{li2017market}, peer effects \citep{Chakraborty-etal-2022}, and 
automakers' responses to environmental policies 
\citep{Aghion-etal-2016, gillingham2021designing, 
lohawala2023roadblock}.}

Turning to the second dimension--the scale and representativeness of data--we 
observe that much of the existing research on EV purchase intent relies on 
single-year surveys, non-representative samples, or relatively small 
datasets. For example, \citet{tiwari2020public}, \citet{hu2023policy}, and 
\citet{mamkhezri2025public} analyzed 1,800 responses from UK residents, 807 
responses from China, and 1,500 responses from the US, 
respectively.\footnote{Although some recent studies (e.g., 
\citet{pani2023decoding} and \citet{ruan2022public}) use large-scale 
online sentiment analysis to track the evolution of public perceptions of 
EVs, they do not examine individual-level predictors of it.} 
\citet{singh2020review} provide a meta-analysis of 211 peer-reviewed research 
articles between 2009\textendash{}2019, identifying key factors influencing 
public opinion of EVs. We overcome the limitations associated with small data 
by using three nationally representative US opinion polls, with substantially 
larger samples: 11,052 in 2021; 7,173 in 2022; and 7,201 in 2023. This allows 
for greater statistical power and more robust generalizability of our findings

Finally, along the third dimension--use of statistical models to analyze EV 
adoption--we note that most studies have relied on linear regression to 
assess the effects of personal beliefs, preferences \citep{Carley-etal-2013, 
Krause-etal-2013} and positive media coverage \citep{Scherrer-2023}. Some 
studies have employed binary probit and logit models to examine motivational 
factors such as reputation \citep{Buhmann-Criado-2023};  while others have 
used mixed logit models to analyze attitudes toward various EV policy 
incentives, employment changes, and electricity costs 
\citep{mamkhezri2025public}. However, in much of this literature, covariate 
effects surprisingly remains unreported, limiting the interpretability of 
the estimated relationships. Beyond these approaches, structural equation 
models have also been used to examine adoption barriers 
\citep{tiwari2020public}, peer effects \citep{zhao2022determines}, and policy 
incentives \citep{hu2023policy}; while topic modeling and sentiment analysis 
have been applied to track temporal shifts in online sentiment toward EVs 
\citep{ruan2022public}. Nonetheless, all the above mentioned modeling 
frameworks cannot capture the heterogeneity in covariate effects across the 
full distribution of intent, which is necessary for a careful and deeper 
understanding of EV adoption.

We employ a Bayesian approach and contribute to the modeling of EV adoption 
literature in two key ways. First, we estimate an ordinal 
probit model using Markov chain Monte Carlo (MCMC) methods to assess how 
our four factors and demographic characteristics relate to an ordinal measure 
of EV purchase intent. We report covariate effects on the average 
probabilities of each response category, an 
aspect that has received limited attention in prior studies. Second, and more 
importantly, we apply ordinal quantile analysis to examine the 
full spectrum of EV purchase intent, addressing the previously unexplored 
question of whether the determinants of EV adoption vary across different 
levels of intent--a question that cannot be answered using conditional mean 
models (e.g., ordinal probit). We note that this is the first application of 
ordinal quantile modeling in the EV adoption literature, enabling us to 
capture heterogeneity in covariate effects both across the full distribution 
of the intent and across the observed ordinal outcome categories (ranging 
from ``not at all likely" to ``very likely" to purchase).

The results from ordinal probit models indicate that greater exposure to 
EV-related information, stronger perceptions of EVs' environmental benefits, 
and higher confidence in future charging infrastructure development are 
positively associated with EV purchase intent, whereas perceptions of 
stronger government climate action are negatively associated with intent 
across samples. These relationships persist after controlling for 
standard demographic characteristics: younger, urban, more educated, and 
higher-income individuals are more likely to express interest in EVs, while 
women, US-born, and Black respondents report lower intent. The results from 
ordinal quantile analysis uncover substantial heterogeneity in covariate 
effects across the distribution of EV purchase intent. This approach provides 
richer insight into how different factors shape adoption intent. For example, 
it shows that confidence in infrastructure development and belief in 
environmental benefits are not only associated with higher intent among 
likely adopters, but also with a lower probability of reporting no interest 
among more skeptical respondents. Finally, we identify notable gaps between 
the importance of these factors and their prevalence: both infrastructure 
confidence and information exposure remain low despite their strong 
association with purchase intent, highlighting opportunities for targeted 
outreach and investment.

We conclude this section by motivating our choice of a Bayesian approach 
for estimating the ordinal models. In non-linear models such as the ordinal 
probit, regression coefficients do not directly correspond to covariate 
effects--defined as changes in outcome probabilities--and these effects must 
therefore be computed separately for each outcome category. Proper estimation 
of covariate effects requires accounting for uncertainty in both observed 
variables and model parameters. But the latter is not captured in the 
Classical framework resulting in bias in covariate effects
\citep{Jeliazkov-Vossmeyer-2018}. In contrast, the Bayesian framework 
provides a full posterior distribution over the parameters and allows 
uncertainty to be systematically incorporated when computing covariate 
effects. Moreover, our adoption of Bayesian methods is driven by 
methodological necessity: to the best of our knowledge, frequentist 
estimation procedures for ordinal quantile models are not currently 
available, making Bayesian estimation essential in this context.

The remainder of the article is structured as follows. Section~\ref{sec:Data} 
outlines the data and provides a preliminary investigation. 
Section~\ref{sec:Model} introduces the ordinal probit and ordinal quantile 
models, along with their MCMC algorithms for estimation.  In 
Section~\ref{sec:Results}, we present the results from our models and in 
Section~\ref{sec:Dicussion} discuss policy implications. Finally, 
Section~\ref{sec:Conclusion} offers concluding remarks.

%------------------------------------------------------------------------------
\section{Data}\label{sec:Data}
%------------------------------------------------------------------------------
The study draws on survey data from three opinion polls carried out by the 
Pew Research Center over a three-year span. The first survey was 
conducted during April 20-21, 2021 on a sample of 13,749 adults, the second 
during May 2-8, 2022 on a sample of 10,282 adults, and the third during May 
30-June 4, 2023 on a sample of 10,329 adults. After removing all missing 
observations and cleaning the data, we have a sample of 11,052, 7,173, and 
7,201 observations available for analysis\footnote{All survey data and 
additional information on the surveys used 
in this paper are available for download from the Pew Research Center dataset 
portal (and the corresponding author's webpage).}. 
Table~\ref{Table:PewDataSummary} 
shows a descriptive summary of the independent and dependent variables. 
Specifically, it presents the mean and standard deviation of the continuous 
variables, and count and percentage for the categorical variables.

The data utilized offers at least three distinct advantages over existing 
studies that use survey data to gauge EV purchase intent. First, they are 
nationally representative and span multiple years. As a result, the 
estimates we generate are more likely to be robust to cross-sectional 
heterogeneity. Second, the availability of 
ordinal responses to likelihood of 
purchasing EV enables us to model public preference regarding EV 
purchase in much greater detail as compared to studies that use binary 
responses, such as \citep{Buhmann-Criado-2023}. Third, the first and second 
surveys correspond to the Covid-19 period, 
while the third survey was conducted in the post-pandemic phase. Therefore, 
if there was an increase in interest for sustainable products like electric 
vehicles during the pandemic, this should be evident in the data summary, 
with a higher percentage of respondents expressing a preference for buying 
electric vehicles during the pandemic phase.

%------------------------------ Table 1 --------------------------------------
\begin{table}[!h]
\centering \footnotesize \setlength{\tabcolsep}{3pt} 
\setlength{\extrarowheight}{1.4pt}
\setlength\arrayrulewidth{1pt}\caption{Descriptive summary of the 
variables for three Pew Research Center surveys: April 2021; May, 
2022; and May-Jun, 2023. Mean (\textsc{mean}) and standard 
deviations (\textsc{std}) are reported for variables treated as continuous, 
while count (\textsc{count}) and percentages (\textsc{per}) are reported for 
categorical variables.}
\begin{tabular}{lp{6cm}r rrr rrr r}
\toprule
&& \multicolumn{2}{c}{April 2021} && 
\multicolumn{2}{c}{May 2022}
&& \multicolumn{2}{c}{May--June 2023} \\
\cmidrule{3-4} \cmidrule{6-7}  \cmidrule{9-10}
\textsc{variable}      &     & \textsc{mean}     & \textsc{std}
&     & \textsc{mean}     & \textsc{std}
&     & \textsc{mean}     & \textsc{std}                            \\
\midrule
%-----------------------------------------------------------------------------
\textsc{Income/10000 (USD)}
&     & 7.00   &  3.33   &&  8.00  &   3.23   &&  8.00 &   3.24   \\
%\midrule
&   & \textsc{count} & \textsc{per} &&  \textsc{count} & 
\textsc{per} &&  \textsc{count} & \textsc{per} \\
\midrule
%-------------------------------------------------------
{EV Info}    & {$\circ$ Read or heard a lot}
&    4170  &  37.73  &&  $..$   &  $..$  &&  $..$   &  $..$   \\
          & {$\circ$ Little or nothing at all}
&    6882  &  62.27  &&  $..$   &  $..$  &&  $..$   &  $..$   \\
\midrule
%-------------------------------------------------------
{Env Better}    & {$\circ$ EVs are environmentally `better' than 
gas-powered vehicles}
&    7944  &  71.88  &&  $..$   &  $..$  &&  $..$  &   $..$   \\
& {$\circ$ About the same or worse}
&    3108  &  28.12  &&  $..$   &  $..$  &&  $..$  &   $..$   \\
\midrule
%-------------------------------------------------------
{EV Infra}    & {$\circ$ US will build charging stations and 
	infrastructure: extremely or very confident}
&    $..$  &  $..$  &&  $..$   &  $..$  &&  1263   &   17.54   \\
& {$\circ$ Somewhat confident, not too confident, or not at all confident}
&    $..$  &  $..$  &&  $..$   &  $..$  &&  5938   &   82.46   \\
\midrule
%-------------------------------------------------------
{EV Owner}    & {Owns electric or hybrid vehicle}
&    1039  &  9.40  &&  769   &  10.72  &&  812   &   11.28   \\
\midrule
%-------------------------------------------------------
{GACC}    & {$\circ$ Federal government is doing `too much' to reduce 
the effects of climate change}
&    1962  &  17.75  &&  1684   &  23.48  &&  1736  &   24.11   \\
& {$\circ$ About the right amount}
&    2203  &  19.93  &&  1636   &  22.81  &&  1582  &   21.97   \\
& {$\circ$ Too little}
&    6887  &  62.31  &&  3853   &  53.72  &&  3883  &   53.92   \\
\midrule
%-------------------------------------------------------
{Age $< 50$}    & {Indicator for age below 50}
&    5124  &  46.36  &&  2687   &  37.46  &&  2596  &   36.05   \\
%-------------------------------------------------------
{Female}    & {Indicator for female gender}
&    5848  &  52.91  &&  3819   &  53.24  &&  3815  &   52.98   \\
\midrule
%-------------------------------------------------------
& {$\circ$ Graduate and above}
&    6447  &  58.33  &&  3846   &  53.62  &&  3730  &   51.80   \\
{Education}     & {$\circ$ Some college}
&    3194  &  28.90  &&  2201   &  30.68  &&  2276  &   31.61   \\
& {$\circ$ HS and below}
&    1411  &  12.77  &&  1126   &  15.70  &&  1195  &   16.59   \\
\midrule
%-------------------------------------------------------
{Married}
& {Married or living with a partner}
&   $..$   &  $..$  &&  5692   &  79.35  &&  5640   &   78.32   \\
% & {$\circ$ divorced, separated, or widowed}
% &   249   &  42.28  &&  527   &  35.39  &&  448   &   35.47   \\
\midrule            
%-------------------------------------------------------
{Metropolitan}    & {Lives in a metropolitan area}
&    9817  &  88.83  &&  6260   &  87.87  &&  6286  &   87.29   \\
\midrule
%-------------------------------------------------------
& {$\circ$ Northeast}
&    1784  &  16.14  &&  1100   &  15.34  &&  1084  &   15.05   \\
{Region}     & {$\circ$ Midwest}
&    2444  &  22.11  &&  1613   &  22.49  &&  1586  &   22.02   \\
& {$\circ$ South}
&    4245  &  38.41  &&  2841   &  39.61  &&  2987  &   41.48   \\
& {$\circ$ West}
&    2579  &  23.34  &&  1619   &  22.57  &&  1544  &   21.44   \\
\midrule
%-------------------------------------------------------
{US Born}
& {Indicator for born in the US}
&   9664   &  87.44  &&  6150   &  85.74  &&  6188  &   85.93   \\
\midrule
%--------------------------------------------
& {$\circ$ White}
&   7801   &  70.58  &&  5335   &  74.38  &&  5083  &   70.59   \\
{Race}
& {$\circ$ Black}
&   820    &  7.42   &&  446   &  6.22   &&   726   &   10.08   \\
& {$\circ$ Other races}
&   2431   &  22.00  &&  1392  &  19.41  &&   1392  &   19.33   \\
\midrule
%-------------------------------------------------------
& {$\circ$ Democrat}
&   4201   &  38.01  &&  2258   &  31.48  &&  2268   &  31.50   \\
{Party}
& {$\circ$ Republican}
&   2667   &  24.13  &&  2453   &  34.20  &&  2341   &  32.51   \\
& {$\circ$ Independent \& others}
&   4184   &  37.86  &&  2462   &  34.32  &&  2592   &  36.00   \\
%-------------------------------------------------------
\midrule
& {Very likely}
&   2411   & 21.82   &&   1526  &  21.27 &&  1397  &   19.40  \\
{Purchase EV}      & {Somewhat likely}
&   2819   & 25.51   &&   1936  &  26.99 &&  1815  &   25.20  \\
&  {Not too likely}
&   3381   & 30.59   &&   1787  &  24.91 &&  1736  &   24.11  \\
      & {Not at all likely}
&   2441   & 22.09   &&   1924  &  26.82 &&  2253  &   31.29  \\
%-------------------------------------------
\bottomrule
\end{tabular}
\label{Table:PewDataSummary}
\end{table}
%------------------------------------------------------------------------------

We begin by discussing the four independent variables that are of primary 
interest in this article. Two such variables, `\texttt{EV	Info}' and 
`\texttt{Env Better}', are specific 
to the April 2021 survey. The variable `\texttt{EV Info}' measures how much a 
respondent has read or heard about EVs. It is coded as 1 if the respondent 
has read or heard a lot about EVs, and 0 if they have read or heard little or 
nothing. As shown in Table~\ref{Table:PewDataSummary}, about 38\% of 
respondents have read or heard a lot about EVs. The variable `\texttt{Env 
Better}' reflects perception about the environmental impact of EVs as 
compared to gas-powered vehicles. This indicator variable takes the value 1 
if the respondent believes that EVs are better for the environment than 
gas-powered vehicles, and 0 otherwise (worse for the environment or about the 
same). According to Table~\ref{Table:PewDataSummary}, approximately 72\% of 
respondents view EVs as being environmentally superior to gas-powered 
vehicles.

The third variable of interest `\texttt{EV Infra}' is only 
available in the May-June 2023 survey. This indicator variable is 
coded as 1 if the respondent is extremely or very confident that the US will 
build the necessary charging stations and infrastructure to support large 
number of EVs on the roads, and 0 otherwise (somewhat confident, not too 
confident, or not at all confident). As seen in 
Table~\ref{Table:PewDataSummary}, about 82\% of respondents 
express a negative perception towards the government for creating the 
infrastructure needed to support the adoption of EVs. The fourth variable of 
interest `\texttt{GACC}' reflects respondents' perceptions of government 
actions (GA) to mitigate global climate change (CC) and is available in all 
three surveys. This variable is divided into three categories: 
`too much', `about the right amount', and `too little' (with `too little' 
serving as the reference category). Table~\ref{Table:PewDataSummary} 
indicates that in each sample, more than half of the respondents feel that 
government efforts to combat global climate 
change are insufficient.

Another variable, which is an important predictor for EV purchase, is 
\texttt{EV Owner}, indicating whether the respondent currently owns an 
electric or hybrid vehicle and taking the value 1 if true and 0 otherwise. As 
shown in Table~\ref{Table:PewDataSummary}, the percentages of respondents who 
are current owners of an electric or hybrid vehicle are 9.40\%, 10.72\%, and 
11.28\% across the three survey datasets. We also incorporate a broad set of 
demographic variables as covariates in our regression models. These variables 
(except marital status) are common across all surveys  and include: income, 
age, gender, education, marital status (not 
available in the April 2021 survey), metropolitan area, region of residence, 
birth country, racial background, and party affiliation. We now examine the 
construction of these variables in more detail.

Pew surveys record income as belonging to one of nine categories: 
$0-30k, 30k-40k, 40k-50k, 50k-60k, 60k-70k, 70k-80k, 80k-90k, 90k-100k$, and 
$>100k$, where $k$ denotes a thousand dollars. To simplify, we use the 
mid-point of each income range and assign an imputed value of \$20,000 for 
the first category and \$110,000 for the last category. Age is recorded into 
four categories (18-29, 30-49, 50-64, and $\ge 65$ years), but for 
convenience we recode it to a binary indicator taking the value 1 if the
respondent is below 50, and 0 otherwise. As shown in 
Table~\ref{Table:PewDataSummary}, 46.36\% of respondents in the first dataset 
belong to the younger age group, with this percentage being 
approximately 10\% lower in the following two datasets. Gender is 
represented by a binary indicator `\texttt{Female}', where 1 indicates female 
and 0 indicates male. We see that more than 50\% respondents across all 
three datasets are female. Education is 
classified into three categories: `\texttt{Graduate \& Above}', 
`\texttt{Some College}', and `\texttt{High School (HS) and Below}'; the 
latter is used as the reference category. As shown in 
Table~\ref{Table:PewDataSummary}, the majority of respondents are 
graduates or have higher education, followed by those with some college 
education. Therefore, respondents across the three samples are generally 
well-educated.

Information on marital status is available only in the May 2022 and 
May$-$June 2023 surveys. We create an indicator variable `\texttt{Married}' 
that takes value 1 if the respondent is married or living with a 
partner, and 0 if they are divorced, separated, widowed, or never been 
married. As per Table~\ref{Table:PewDataSummary}, more than three-quarters of 
respondents are married or living with a partner. Two variables reflect a 
respondent's residential background. The `\texttt{Metropolitan}' variable is 
an indicator that equals 1 if the respondent lives in a metropolitan area, 
and 0 otherwise. Across the three datasets, more than four-fifths of the 
respondents are located in metropolitan areas. Additionally, we include 
regional indicators for `\texttt{Northeast}', `\texttt{Midwest}', 
`\texttt{South}', and `\texttt{West}'; with \texttt{Northeast} serving as the 
reference category in our regression models. As shown in 
Table~\ref{Table:PewDataSummary}, the largest proportion of respondents 
(about 40\%) reside in the South, followed by 
approximately 22\% in both the Midwest and West. The smallest fraction, 
about 15\%, lives in the Northeast.

The variable `\texttt{US Born}' indicates whether the respondent was born in 
the US or is an immigrant. According to Table~\ref{Table:PewDataSummary}, a 
predominant fraction (about 86\%) of the sample across three datasets were 
born in the US. Our demographic variable list also includes race indicators 
for `\texttt{White}' (reference category), `\texttt{Black}', and 
`\texttt{Other Races}'. As expected, the largest proportion of respondents 
(over 70\%) are White, followed by  about 20\% identifying as belonging to 
other races. The smallest group, about 8\%, are Black. Finally, party 
affiliation is captured with indicator variables for `\texttt{Democrat}', 
`\texttt{Republican}' (reference category), and `\texttt{Independent \& 
Others}'. Respondents are roughly equally distributed across these three 
political categories.

%---------------------------  Figure  ---------------------------------------
\begin{figure*}[!htb]
	\centerline{
		\mbox{\includegraphics[width=7.25in, height=8.25in]{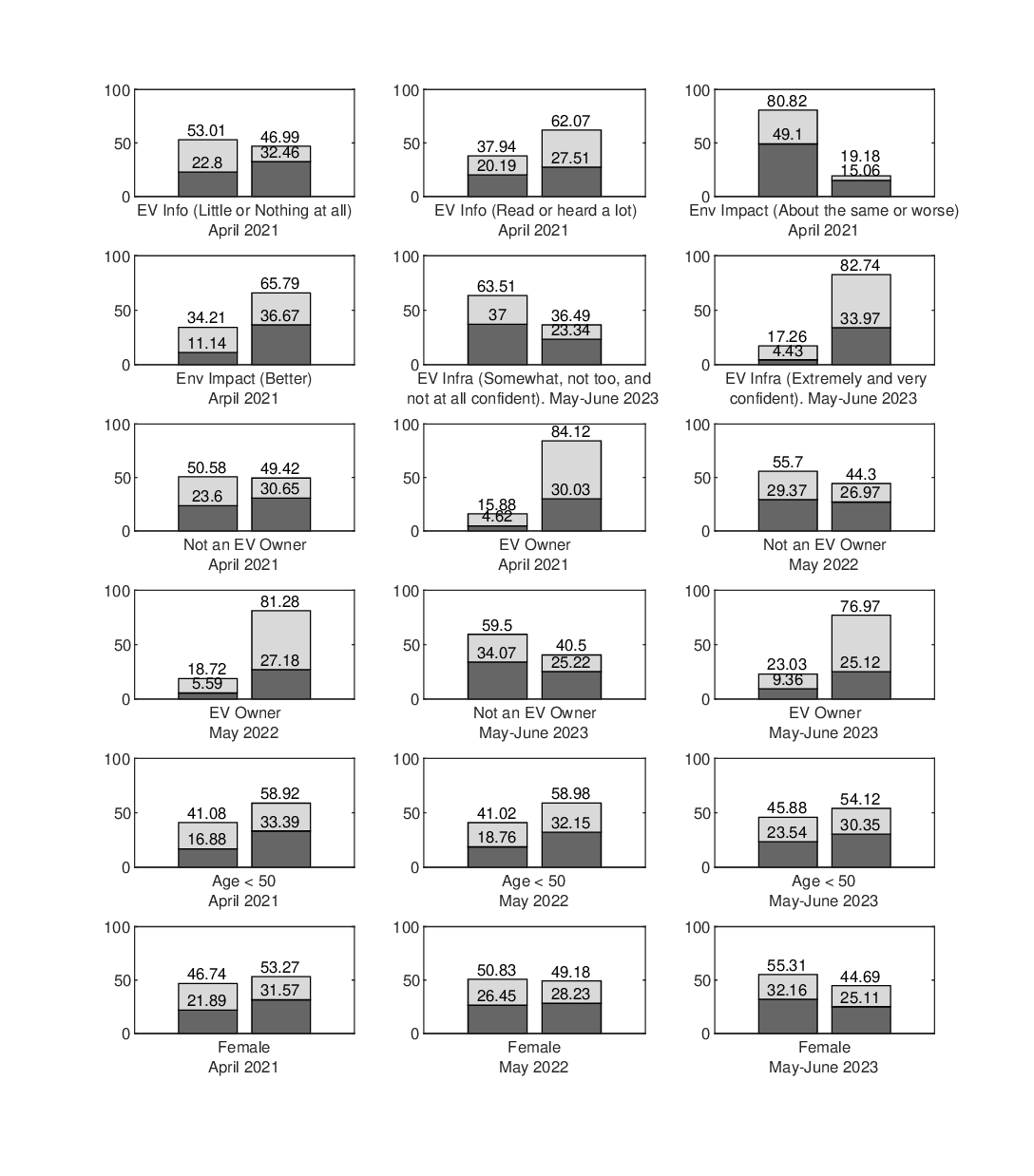}}
	}
	\vspace{-3pc}
	\caption{Stacked bars displaying the percentage of observations 
	corresponding to the four opinion categories for a selected list of 
	covariates. In each panel, the first stacked bar 
	shows the percentage of observations for `not at all likely' and the 
	cumulative percentage for `not at all likely' and `not too likely'. The 
	second stacked bar shows the percentage of observations for `somewhat 
	likely' and the cumulative percentage for `somewhat likely' and 
	`very likely'.}
	\label{fig:StackedBars}
\end{figure*}
%-------------------------------------------------------------------------------
 
As noted earlier, our objective is to analyze the factors that influence 
individuals' decision to purchase an EV. However, a continuous measure of 
support or opposition is not directly observable. The data is limited to the 
respondent's answer to the question: ``The next time you purchase a 
vehicle, how likely are you to seriously consider purchasing an electric 
vehicle?'' Response to this question serves as the dependent variable in our 
models and has four categories: `very likely', `somewhat likely', `not too 
likely', and `not at all likely'. The surveys also include a fifth category 
:`I do not expect to purchase a vehicle'; we exclude responses in this 
category from our analysis to simplify the framework and adhere to ordinal 
modeling. Table~\ref{Table:PewDataSummary} presents the count and percentage 
of each response category in the three datasets. We see from 
Table~\ref{Table:PewDataSummary}, 47.33\% of the 
respondents in the April 2021 survey and 48.26\% of the respondents in the 
May 2022 survey are either `very likely' or `somewhat 
likely' to purchasing an EV. In comparison, the corresponding percentage in 
the May-June 2023 survey is 44.60\%. Recall, the first two surveys were 
conducted during the Covid-19 period, while the third survey took place in 
the post-Covid period. Based on these percentages, we do not observe a 
significant decline in preference for purchasing EVs in the post-pandemic 
phase. Overall, our data summary clearly shows that public attitude towards 
purchasing an EV is generally positive.

Further insights into the association between the dependent variable and  
selected regressors from the three survey datasets can be gained from
Figure~\ref{fig:StackedBars}. In each panel, the bars represent two 
`opinion' categories: `not likely' and `likely'. The first stacked bar
displays the percentage of responses for `not at all likely' (shown in dark 
shade) and the cumulative percentage for both `not at all likely' and `not 
too likely' (sum of dark and light shades). The second stacked bar 
illustrates the percentage of responses for `somewhat likely' (in dark shade) 
and the cumulative percentage for `somewhat likely' and `very likely' (sum of 
dark and light shades).

The first two panels of Figure~\ref{fig:StackedBars} highlight the 
influence of information on consumers' decisions to purchase EVs. 
In the first panel, among respondents who have `read or heard little or 
nothing at all' about EVs, only 46.99\% are likely to consider 
purchasing one. In contrast, the second panel reveals a much higher 
likelihood of purchase---62.07\%---among those who have `read or heard a 
lot' about EVs. A more striking difference emerges when examining perceptions 
of the environmental impact of vehicles. As shown in the third panel, just 
19.18\% of respondents who view EVs as `about the same or worse' than 
gas-powered vehicles are likely to purchase one. However, as shown in the 
fourth panel, the percentage jumps to 65.79\% for those who believe EVs have 
a lower environmental impact. A similar pattern is evident, from panels five 
and six, when considering respondents' views on the development of 
charging stations and infrastructure to support large number of EVs on the 
road. Among those who lack confidence that the government will build the 
necessary facilities, only 36.49\% are likely to buy an EV. 
In stark contrast, a much higher 82.74\% of respondents who believe the 
government will build the required facilities for EVs are likely to make a 
purchase. These relationships highlight how knowledge and environmental 
impact of EVs, and confidence in supporting infrastructure play pivotal roles 
in shaping consumer purchase decisions.

The remainder of panels in Figure~\ref{fig:StackedBars} (i.e., panels 
seven to eighteen) exhibit how factors such as ownership of an EV or hybrid 
vehicle, age, and gender affect the decision to purchase an EV. The stacked 
bars from panels seven to twelve clearly show that owning an EV or hybrid 
vehicle (as opposed to not) increases the likelihood of purchasing an EV 
across all survey datasets. For instance, 84.12\% of EV or hybrid owners are 
likely to purchase an EV as compared to only 49.42\% (a difference of 34.7\%) 
in the April 2021 survey data. For the same comparison, a difference of 
36.98\% ($81.28\% - 44.3\%$) and 36.47\% ($76.97\% - 40.5\%$) is noted in the 
May 2022 and May-June 2023 survey data. Age also plays a role, with a 
majority of respondents under 50 being likely to purchase an EV across all 
three datasets. With respect to gender, the likelihood of purchasing an EV 
among females is 53.27\% in the April 2021 dataset, decreases to 49.18\% 
in May 2022, and further drops to 44.69\% in the May-June 2023 dataset.

%------------------------------------------------------------------------------
\section{The Modeling Framework}\label{sec:Model}
%------------------------------------------------------------------------------
In this section, we briefly describe the ordinal probit model followed by the 
ordinal quantile model. For both the models, we outline the Bayesian 
estimation algorithm using Markov chain Monte Carlo (MCMC) methods and 
explain the computation of covariate effects.

\subsection{The Ordinal Probit Model}\label{sec:OrdProbit}

\emph{Ordinal} data models \citep{Johnson-Albert-2000}, also known as ordered 
choice models, allows fitting a regression model to an ordinal dependent 
(outcome or response) variable, typically denoted by $y$, as a function of 
the covariates. These models are grounded in the random 
utility framework from economics and explained, amongst others, in 
\citet{Jeliazkov-Rahman-2012} and 
\citet{Batham-etal-2023}. In a typical setting, the response 
variable $y$ has several categories and each category is assigned a score 
(value or number) which is inherently ordered or ranked, without any cardinal 
interpretation. For example, in our study, responses ($y$) to the 
question about purchasing an EV are coded as follows: 1 for 
`not at all likely', 2 for `not too likely', 3 for `somewhat likely', and 4 
for `very likely'. Here, a score of 2 as compared to 1 implies more 
inclination to buy an electric vehicle, but we cannot interpret a score of 2 
as twice the inclination compared to a score of 1. Similarly, the difference 
in inclination between the scores 2 and 1 is not the same as that between 4 
and 3.

The ordinal regression model, following \citet{Albert-Chib-1993}, can be 
written in terms of a continuous latent variable $z_{i}$ as follows:
%--------------------------
\begin{equation}
	z_{i} = x'_{i} \beta  + \varepsilon_{i}, \hspace{0.75in} \forall \; i=1, 
	\cdots, n,
	\label{eq:OrdCont}
\end{equation}
%--------------------------
where $x_{i}$ is a $k \times 1$ covariate vector including a column of ones, 
$\beta$ is a $k \times 1$ parameter vector, and $n$ is the number of 
observations. We assume the error $\varepsilon$ is \textit{independent 
	and identically distributed} (\textit{iid}) as a 
standard normal distribution i.e., $\varepsilon_{i} \sim N(0,1)$ for $i=1,2, 
\ldots,n$; which gives rise to the 
\textit{ordinal probit} model. 

\begin{figure*}[!t]
	\centerline{
		\mbox{\includegraphics[width=7.0in, height=2.2in]{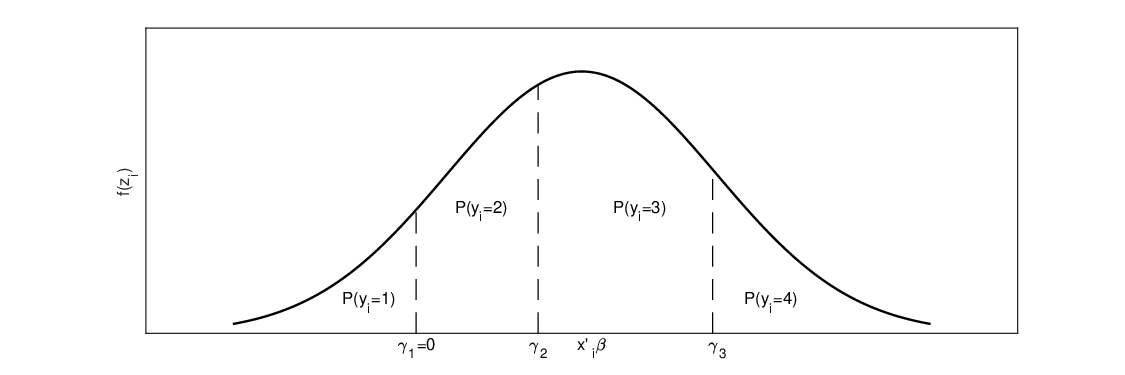}}
	}
	\caption{A pictorial representation of the distribution of the latent 
		variable $z$. The four probabilities P$(y_{i}=1)$, P$(y_{i}=2)$, 
		P$(y_{i}=3)$ and P$(y_{i}=4)$ correspond to the responses: `not at 
		all likely', `not too likely', `somewhat likely', and `very likely', 
		regarding the intention to purchase an electric vehicle ($z_{i}$) for 
		individual $i$ with mean $x'_{i}\beta$.}
	\label{fig:ordinalpdf}
\end{figure*}

In our study, the unobserved variable $z_{i}$ represents an individual's 
intent to purchase EV and is linked to the observed ordinal response $y_{i}$ 
as follows:
%------------------------------------------------------
\begin{equation}
	\gamma_{j-1} < z_{i} \leq \gamma_{j} \; \longleftrightarrow \;
	\emph{$y_{i}$ = j}, \hspace{0.75in}
	\forall \; i=1,\cdots, n; \; j=1,\cdots, J,
	\label{eq:cutpoints}
\end{equation}
%------------------------------------------------------
where $-\infty = \gamma_{0} < \gamma_{1} \cdots < \gamma_{J-1} < \gamma_{J} =
\infty$ are the cut-points (or thresholds) and $J$ denotes the number of 
categories or outcomes of $y$. This relationship between the latent $z$ 
(intent to purchase an EV) and observed $y$ (response to the question about 
purchasing an EV) is illustrated in Figure~\ref{fig:ordinalpdf}. As 
shown, $P(y_{i}=1)$ is the area under $f(z_{i})$ to the left 
of $\gamma_{1}$, $P(y_{i}=2)$ is the area under $f(z_{i})$ between 
$\gamma_{1}$ and $\gamma_{2}$, and so forth. To specify the location and 
scale restrictions, we fix $\gamma_{1}=0$ and $var{(\varepsilon)}=1$, 
respectively \citep{Jeliazkov-etal-2008, Jeliazkov-Rahman-2012}.

%----------------------------  Figure  ---------------------------------------
%-------------------------------------------------------------------------------

Given the responses $y$ = $(y_{1}, \cdots, y_{n})'$ and the covariates, the 
likelihood  function for ordinal probit model can be written as follows: 
%---------------------------------
\begin{equation}
	\begin{split}
		f(y|\beta, \gamma) & = \prod_{i=1}^{n} \prod_{j=1}^{J} \Pr(y_{i} = j 
		| 
		\beta,	\gamma)^{ I(y_{i} = j)} 
		=  \prod_{i=1}^{n}  \prod_{j=1}^{J}
		\bigg[ \Phi(\gamma_{j} - x'_{i}\beta) -
		\Phi(\gamma_{j-1} - x'_{i}\beta)   \bigg]^{ I(y_{i} = j)},
		\label{eq:likelihoodOP}
	\end{split}
\end{equation}
%---------------------------------
where $\Phi(\cdot)$ is the cumulative distribution function (\emph{cdf})
of a standard normal distribution and $I(y_{i}=j)$ is an indicator function,
which equals 1 if the condition within parenthesis is true and 0 otherwise. 
But the ordering constraint $\gamma_{1} = 0 < \gamma_{2} < ... < 
\gamma_{J-1}$ is difficult to satisfy during the MCMC sampling. So, following 
\citet{Albert-Chib-2001}, we apply the transformation $\delta _{j}= \ln 
(\gamma_{j} - \gamma_{j-1})$, for $2 \leq j \leq J-1$, and rewrite the 
likelihood as a function of $\beta$ and $\delta = (\delta_2, \ldots, 
\delta_{J-1})'$, which allows us to make inferences from $f(y|\beta, 
\delta)$. 

To complete the Bayesian framework, we utilize the Bayes theorem to combine 
the complete data likelihood $f(y,z|\beta, \delta)$ with the prior 
distributions: $\beta \sim N\left( \beta_{0}, B_{0} \right)$ and $\delta \sim 
N \left( d_0, D_0 \right)$. The resulting complete data posterior can be 
written as,
%---------------------------
\begin{equation}
	\begin{split}
		\pi \left( \beta, \delta, z | y \right) & \propto f\left( y | \beta,
		\delta, z \right) f ( z | \beta)  \pi (\beta) \pi (\delta) 
		= \left\{ \prod\limits_{i=1}^{n} f\left( y_{i} | \delta, z_i \right) 
		\right\} f_N (z | X\beta,I_n)  \pi (\beta) \pi (\delta),
	\end{split}
	\label{eq:OPAugJointPost}
\end{equation}
%---------------------------
where $f ( y_{i} | \delta, z_i ) = 1 \{ \gamma_{j-1} < z_{i} \le \gamma_j
\}$, with the index $j$ determined by $y_{i}$, and $\pi(\beta)$ and 
$\pi(\delta)$ are the normal priors. Given 
the complete data posterior \eqref{eq:OPAugJointPost}, the conditional 
posteriors can be derived and the model estimated using MCMC sampling as 
outlined in Algorithm~1.

Post estimation, an important construct is to compute the 
covariate effects (or marginal effects) on the outcome probabilities, since, 
unlike linear regression, the coefficients do not give the 
covariate effects. Let the $l$-th 
covariate, ${x_{i,l}}$, take two distinct values, $a$ and $b$, denoted 
as ${x_{i,l}^{a}}$ and ${x_{i,l}^{b}}$, respectively. We partition the 
covariate and parameter vectors as follows: $x_{i}^{a} = (x_{i,l}^{a},
x_{i,-l})$, $x_{i}^{b} = (x_{i,l}^{b}, x_{i,-l})$, and $\beta = 
(\beta_{l}, \beta_{-l})$, where $-l$ denotes all covariates (and 
parameters) except for the $l$-th covariate (and parameter). We evaluate the 
distribution of the difference ${\Pr(y_{i}=j|x_{i,l}^{b}) - 
	\Pr(y_{i}=j|x_{i,l}^{a}) }$ for $1 \le j \le J$, by marginalizing over 
${x_{i,-l}}$ and $(\beta, \delta)$. This enables us to account for 
sampling and parameter uncertainty, and avoid nontrivial biases in estimating 
covariate effects \citep{Jeliazkov-Vossmeyer-2018}. 

%---------------------------
\vspace{10pt} \textbf{Algorithm~1}: MCMC algorithm for estimating ordinal 
probit model.

\noindent\rule{\textwidth}{0.5pt}
\begin{enumerate}
	\setstretch{1} 
	\item Sample $\delta, z | y, \beta$ in one block as follows:
	%--------------------------------
	\begin{enumerate}
		\item Sample $ \delta |y, \beta $ marginally of $z$ by generating 
		$\delta'$ from a random-walk chain $\delta' = \delta + s$, where $s 
		\sim 
		N(0_{J-2}, \iota^{2} \hat{D})$, $\iota$ is a tuning parameter, 
		$\hat{D} = 
		\left. -\left[ \frac{ \partial^2 \ln f( y |\beta, \delta)}{\partial 
			\delta \partial \delta^{T}} \right]^{-1} \right|_{\delta = 
			\hat{\delta}}$, and $\hat{\delta} = \arg \max_\delta \ln f (y | 
		\beta, 
		\delta )$. Accept the proposed value $\delta'$ with probability%
		$$\alpha _{MH}( \delta , \delta') = \min \left\{ 1, \frac{ f( y | 
			\beta,\delta') \pi (\delta')}
		{f( y | \beta, \delta ) \pi(\delta)} \right\},$$
		otherwise repeat the current value $\delta$.%
		
		\item Sample $z_{i} | y, \beta ,\delta \sim TN_{\left( \gamma
			_{j-1},\gamma _{j} \right) } \left( x_{i}' \beta, 1 \right) $ for
		$i = 1, \ldots, n$, where $TN_{( \gamma_{j-1},\gamma _{j} )}$ 
		denotes a truncated normal distribution constrained between 
		$\gamma_{j-1}$ and $\gamma_{j}$; and $\gamma $ is derived from the 
		one-to-one mapping relating $\gamma$ and $\delta$.
	\end{enumerate}
	%--------------------------------
	\item Sample $\beta | z \sim N( \tilde{\beta}, \tilde{B} )$ where
	$\tilde{B}^{-1} = ({B}_{0}^{-1} + X'X)$, and $\tilde{\beta} = 
	\tilde{B} (B_{0}^{-1} \beta_0 + X'z)$.
\end{enumerate}
\vspace{-10pt}
\noindent\rule{\textwidth}{0.5pt} 
%---------------------------
\smallskip

To obtain draws from the distribution ${\Pr(y_{i}=j|x_{i,l}^{b}) - 
\Pr(y_{i}=j|x_{i,l}^{a})}$, we use the method of composition
\citep{Jeliazkov-etal-2008}. This involves 
selecting an individual, extracting the corresponding covariate values, 
drawing $(\beta, \delta)$ from their posterior distributions, and evaluating 
the difference ${\Pr(y_{i}=j|x_{i,l}^{b},x_{i,-l},\beta,\delta) - 
\Pr(y_{i}=j|x_{i,l}^{a},x_{i,-l},\beta,\delta)}$, where $
\Pr(y_{i}=j|x_{i}^{q},\beta,\delta)  = 
\Phi(\gamma_{j} - x_{i,l}^{q} \, \beta_{l} - x'_{i,-l} \,
\beta_{-l}) - \Phi(\gamma_{j-1} - x_{i,l}^{q} \,
\beta_{l} - x'_{i,-l} \, \beta_{-l}),$
%----------------------------------
for $q=b, a$, and $1 \le j \le J$. The process is repeated for all 
individuals and MCMC draws. The average covariate effect (ACE) for outcome 
$j$ is then computed as the mean of pointwise probability differences:
%-------------------------------------
\begin{equation}\label{eq:ACEOR1}
	ACE_{y=j} \approx \frac{1}{M} \frac{1}{n} \sum_{m=1}^{M} 
	\sum_{i=1}^{n} 
	\Big[ \Pr(y_{i}=j|x_{i,l}^{b},x_{i,-l},\beta^{(m)},\delta^{(m)}) -
	\Pr(y_{i}=j|x_{i,l}^{a},x_{i,-l},\beta^{(m)},\delta^{(m)} ) 
	\Big],
\end{equation}
%----------------------------------
where $(\beta^{(m)},\delta^{(m)})$ denotes the MCMC draws, and $M$ is the 
number of post burn-in MCMC draws.

\subsection{The Ordinal Quantile Model}\label{sec:OrdQuantile}
%-----------------------------------------------------------------------------
The parameter estimates from the ordinal probit model characterize the 
expected value of the latent response variable and the average probabilities 
of outcomes conditional on the covariates. However, policy-relevant questions 
often require examining how covariates affect specific parts of the outcome 
distribution, particularly the lower and upper extremes. For instance, having 
confidence that public charging infrastructure will continue to expand---one 
of our key covariates---may have less effect on individuals with low interest 
in EVs. However, among those who are moderately interested, increased 
confidence in future infrastructure may strengthen intent, reassuring them 
that charging access will not be a barrier and moving them from ``somewhat 
likely'' to ``very likely'' buyers. To capture such heterogeneity in 
covariate effects across different levels of purchase intent, we employ 
ordinal quantile regression.

The ordinal quantile regression model, proposed by \citet{Rahman-2016}, can 
be written in terms of a latent variable $z_{i}$ as follows:
%--------------------------
\begin{equation}
	\begin{aligned}
		z_{i} & = x'_{i} \beta_{p}  + \varepsilon_{i}, && \qquad \forall \; 
		i=1, \cdots, n, \\
		\gamma_{p,j-1} & < z_{i} \le \gamma_{p,j} \; \longleftrightarrow \;
		\emph{$y_{i}$ = j}, && \qquad 
		\forall \; i=1,\cdots, n; \; j=1,\cdots, J,
	\end{aligned}
	\label{eq:OrdQuantModel}
\end{equation}
%--------------------------
where the regression coefficient $\beta_{p}$ and cut-point vector 
$\gamma_{p}$ are quantile dependent as indicated by the subscript $p$, and 
$\varepsilon_{i} \overset{iid}{\sim} AL(0,1,p)$ for $i=1,2, \ldots,n$; where 
AL denotes an  asymmetric Laplace distribution \citep{Yu-Zhang-2005}. The AL 
distribution is essential to create a working likelihood function, 
and yields the following expression, 
%--------------------------
\begin{equation}
	\begin{split}
		f(y|\beta_{p},\gamma_{p})
%		& =  \prod_{i=1}^{n} \prod_{j=1}^{J} P(y_{i} = j | 
%		\beta_{p},\gamma_{p})^{ I(y_{i} 
%			= j)}  \\
		& =  \prod_{i=1}^{n}  \prod_{j=1}^{J}
		\bigg[ F_{AL}(\gamma_{p,j} - x'_{i}\beta_{p}) -
		F_{AL}(\gamma_{p,j-1} - x'_{i}\beta_{p})   \bigg]^{ I(y_{i} = j)},
		\label{eq:OrdQuantLike}
	\end{split}
\end{equation}
%--------------------------
where $F_{AL}(\cdot)$ denotes the 
\emph{cdf} of an AL distribution and $I(y_{i}=j)$ is an indicator function.

At this stage, three key issues need attention. First, the 
ordering constraint of cut-points during sampling is addressed 
using the transformation, $\delta_{p,j} = \ln ( \gamma_{p,j} - 
\gamma_{p,j-1} )$ for $2 \le j \le J-1$. Second, we set
$\gamma_{p,1} = 0$ to anchor the location and variance is automatically fixed 
for any quantile to meet the scale restriction. Third, the AL distribution is 
not convenient for MCMC sampling as it does not yield tractable conditional 
posteriors. So, following \citet{Kozumi-Kobayashi-2011}, we 
utilize the normal-exponential mixture formulation: 
$\varepsilon_{i} = \theta w_{i} + \tau \sqrt{w_{i}} \, 
u_{i}$, where $\theta = (1-2p)/(p(1-p))$ and $\tau = 
\sqrt{2/(p(1-p))}$. The variable $w_{i}$ follows an exponential distribution, 
$w_{i}  \sim\mathcal{E}(1)$, and is independent of $u_{i}$, which follows a 
normal distribution, $u_{i} \sim N(0,1)$. Based on this formulation, 
$z_{i}|\beta_{p},w_{i} \sim N(x'_{i}\beta_{p} + \theta w_{i}, \tau^{2} 
w_{i})$ for $i=1,\ldots,n$; enabling us to leverage the properties of normal 
distribution for an efficient MCMC algorithm.

By Bayes theorem, we combine the complete data likelihood 
$f(y,z|\beta_{p},\delta_{p},w)$ with the normal priors: 
$\beta_{p} \sim N(\beta_{p0}, B_{p0})$, $\delta_{p} \sim N(\delta_{p0}, 
D_{p0})$, to arrive at, 
%--------------------------
\begin{equation}
	\begin{split}
		&\pi(z,\beta_{p}, \delta_{p},w|y)
		 \propto f(y|z,\beta_{p},\delta_{p},w) \; f(z | \beta_{p}, w)
		\; \pi(w) \; \pi(\beta_{p})  \; \pi(\delta_{p}),  \\
		%-----------------
		& \propto \Big\{ \prod_{i=1}^{n} f(y_{i}|z_{i},\delta_{p}) \Big\}
		\; f(z | \beta_{p},w) \; \pi(w) \; \pi(\beta_{p})
		\; \pi(\delta_{p}), \\
		%-----------------
		& \propto  \prod_{i=1}^{n} \bigg\{ \prod_{j=1}^{J}
		1\{\gamma_{p,j-1} < z_{i} < \gamma_{p,j} \} \;
		N(z_{i}|x'_{i}\beta_{p} + \theta w_{i}, \tau^{2} w_{i})
		\; \mathcal{E}(w_{i}|1) \bigg\} \pi(\beta_{p}) \pi(\delta_{p}),
		%-----------------
		%& \qquad \times \; N(\beta_{p}|\beta_{p0}, B_{p0})
		%\; N(\delta_{p}|\delta_{p0}, D_{p0}).
	\end{split}
	\label{eq:JointPostORI}
\end{equation}
%--------------------------
where $\pi(\beta_{p})$ and $\pi(\delta_{p})$ denote the normal priors.
In the first line, we use the factorization $f(y,z|\beta_{p},\delta_{p},w) = 
f(y|z,\beta_{p},\delta_{p},w) \times f(z|\beta_{p},w)$ and note that $z$ is 
independent of $\delta_{p}$. In the second line, we exploit the independence 
of $y_{i}$ from $(\beta_{p},w)$ given $(z_{i},\delta_{p})$, which follows 
from the second line of equation~\eqref{eq:OrdQuantModel}, where $y_{i}$
given $(z_{i}, \delta_{p})$ is determined with probability 1. In the third
and fourth lines, we specify the conditional distribution of the latent 
variable $z$ and the prior distribution on the parameters 
$(\beta_{p},\delta_{p})$.

%-------------------------------------------------------------------------------
%\begin{table*}[!t]
%\begin{algorithm}[Sampling in $\mathrm{OR_{I}}$ model] \label{alg:algorithm1}

\vspace{10pt} \textbf{Algorithm~2}: MCMC algorithm for estimating ordinal 
quantile model.

\noindent\rule{\textwidth}{0.5pt} 
\begin{enumerate}
	\item  Sample $\beta_{p}| z,w$ $\sim$  $N(\tilde{\beta}_{p}, 
	\tilde{B}_{p})$, where,
	$$\tilde{B}^{-1}_{p} = \bigg(\sum_{i=1}^{n}
	\frac{x_{i} x'_{i}}{\tau^{2} w_{i}} + B_{p0}^{-1} \bigg), \quad
	\mbox{and} \quad \tilde{\beta}_{p} = \tilde{B}_{p}\bigg( 
	\sum_{i=1}^{n}
	\frac{x_{i}(z_{i} - \theta w_{i})}{\tau^{2} w_{i}} + B_{p0}^{-1} 
	\beta_{p0}
	\bigg).$$
	\item Sample $w_{i}|\beta_{p}, z_{i}$ $\sim$  $GIG \, (0.5, 
	\tilde{\lambda}_{i},
	\tilde{\eta}) $, for $i=1,\cdots,n$, where,
	$$\tilde{\lambda}_{i} = \Big( \frac{ z_{i} - x'_{i}\beta_{p}}{\tau}
	\Big)^{2}, \quad \mbox{and} \quad \tilde{\eta} = \Big(
	\frac{\theta^{2}}{\tau^{2}} + 2 \Big).$$
	\item  Sample $\delta_{p}|y,\beta_{p}$ marginally of $w$ and $z$, by 
	generating $\delta_{p}'$ from a 
	random-walk chain $\delta'_{p} = \delta_{p} + s$, where $s \sim N(0_{J-2},
	\iota^{2} \hat{D})$, $\iota$ is a tuning parameter and $\hat{D}$
	denotes negative inverse Hessian, obtained by maximizing
	the log-likelihood with respect to $\delta_{p}$. Given the current 
	value of $\delta_{p}$, accept the proposed draw $\delta'_{p}$ with 
	probability,
	%--------------------------------
	\begin{equation*}
		\alpha_{MH}(\delta_{p}, \delta'_{p}) = \min \bigg\{1,
		\frac{ \;f(y|\beta_{p},\delta'_{p}) \;\pi(\beta_{p}, \delta'_{p})}
		{f(y|\beta_{p},\delta_{p}) \;\pi(\beta_{p}, \delta_{p})}
		\bigg\};
	\end{equation*}
	%--------------------------------
	else repeat value $\delta_{p}$. The variance of $s$ may 
	be tuned to achieve desired acceptance rate.
	\item    Sample $z_{i}|y, \beta_{p},\gamma_{p},w$ $\sim$ 
	$TN_{(\gamma_{p,j-1}, \gamma_{p,j})}(x'_{i}\beta_{p} + \theta w_{i}, 
	\tau^{2}w_{i})$ for $i=1,\cdots,n$, where $TN$ denotes a truncated normal 
	distribution and $\gamma_{p}$ is obtained via $\delta_{p}$ using the 
	one-to-one mapping. 
\end{enumerate}
\vspace{-10pt}
\noindent\rule{\textwidth}{0.5pt}

\vspace{1pc}
%\end{algorithm}
%\end{table*}
%%-----------------------------------------------------------------------------------------

The conditional posterior distributions are derived from the complete data 
posterior (i.e., equation~\ref{eq:JointPostORI}) and the 
parameters are sampled as per Algorithm~2. The implementation of this 
algorithm, available in the 
\href{https://cran.r-project.org/web/packages/bqror/}{bqror} package, is 
explained in \citet{Maheshwari-Rahman-2023}. 

After estimating the ordinal quantile models, we use the method of 
composition to compute the covariate effects, such as in 
\citep{Rahman-Vossmeyer-2019} and \citet{Bresson-etal-2021}. Assuming 
a setup similar to Section~\ref{sec:OrdProbit}, our goal is to draw from the 
distribution $\{\Pr(y_{i}=j|x_{i,l}^{b}) - 
\Pr(y_{i}=j|x_{i,l}^{a} ) \}$. So, we randomly select an individual, extract 
their covariates, draw a value $(\beta_{p}, \delta_{p})$ from their posterior 
distributions, and finally evaluate
$\{\Pr(y_{i}=j|x_{i,l}^{b},x_{i,-l},\beta_{p},\delta_{p}) -
\Pr(y_{i}=j|x_{i,l}^{a},x_{i,-l},\beta_{p},\delta_{p} )\}$, where now for the 
quantile model: $\Pr(y_{i}=j|x_{i}^{q},\beta_{p},\delta_{p}) = 
F_{AL}(\gamma_{p,j} - x_{i,l}^{q} \, \beta_{p,l} - x'_{i,-l} \,
\beta_{p,-l}) - F_{AL}(\gamma_{p,j-1} - x_{i,l}^{q} \,
\beta_{p,l} - x'_{i,-l} \, \beta_{p,-l})$,
%----------------------------------
for $q=b, a$ and $1 \le j \le J$. This process is repeated for all 
individuals and MCMC draws, and the average covariate effect for outcome $j$ 
at quantile $p$ is calculated as the mean of difference in pointwise 
probabilities.

%------------------------------------------------------------------------------
\section{Empirical Results}\label{sec:Results}
%------------------------------------------------------------------------------

%In this section, we report the results from the Bayesian estimation of 
%ordinal probit and ordinal quantile models. We also report the marginal 
%likelihood which may be used for model comparison. Morevoer, for each 
%model, we compute the covariate effects for all or some of the significant 
%variables and discuss our findings in the context of existing literature.

%-----------------------------------------------------------------------------
\subsection{Results from Ordinal Probit Models}

We begin by presenting the results from ordinal probit models 
estimated separately for each survey. Table~\ref{Table:EVProbitResults} 
presents the posterior means and standard deviations of the parameters based 
on 10,000 MCMC iterations after a burn-in of 2,500 iterations. We utilize the 
priors $\beta \sim N_{k}(0, I)$ and $\delta \sim N_{J-2}(0, 0.25 \times I)$, 
where $I$ denotes an identity covariance matrix. These priors are relatively 
diffuse, allowing the data to primarily drive the posterior inference. 
Table~\ref{Table:EVProbitResults} also reports the logarithm of marginal 
likelihood \citep{Chib-Jeliazkov-2001, Jeliazkov-etal-2008}, which can be 
used for Bayesian model comparison. The marginal likelihood (or model 
evidence) is the probability of observed data, obtained by integrating the 
likelihood function with respect to prior distribution. It accounts for both 
model fit and complexity, with higher values indicating stronger support for 
a model\footnote{In Table~2, we fit an 
	ordinal probit model (say M1) to the 2021 survey data and obtain 
	logarithm of marginal likelihood as $\ln f(y|M1) = -12415$. Suppose, we 
	now estimate a reduced model (denoted M0) that excludes the regional 
	indicators, for	which the logarithm of marginal likeilhood is $\ln 
	f(y|M0) = -12425$. Then the logarithm of Bayes Factor (BF) in favor of 
	M1 relative to M0 is, $\ln(BF)_{10} = - 12415-(-12425) = 10$. This 
	constitutes strong evidence in favor of the model with regional 
	indicators. We note that marginal likelihoods (and hence Bayes factors) 
	are only comparable across models fitted to the same dataset; 
	consequently, comparisons across different survey years are not 
	meaningful, as the models are conditioned on different observed data.}.
The use of marginal likelihood contrasts with  
commonly used classical measures, such as the pseudo-$R^2$, Akaike 
Information Criterion (AIC), and Bayesian Information Criterion (BIC) 
\citep{Lin-Tan-2017, Kumar-etal-2025}.

In ordinal models, coefficient signs indicate the direction of covariate 
effects for the first and last response categories. For example, in 
Table~\ref{Table:EVProbitResults}, `\texttt{EV Info}' has a 
positive coefficient (0.43), so it increases (decreases) the probability 
of last (first) category ``Very likely'' (``Not at all likely''). However, these coefficients do not represent the covariate effects nor 
can we state anything \textit{a-priori} about the  effects on 
intermediate outcomes (``Not too likely'' and ``Somewhat likely''). 
Therefore, Table~\ref{Table:OPCE} provides the covariate effects for all 
variables whose parameters are statistically different from zero at 95\% 
probability level.\footnote{As noted earlier, some variables of interest are 
available in only one year. For variables measured in all three survey years, 
however, the estimated effects are similar across years, suggesting that the 
influences on EV purchase intent are relatively stable over time. This 
stability supports the credibility of estimates for variables observed in a 
single year.}

The combined results from the three surveys reveal consistent 
patterns in factors associated with EV purchase intent. As shown in 
Table~\ref{Table:OPCE}, the covariate effects from the 2021 
survey show that respondents who have read or heard extensively about EVs are 
10.2 (2.2) percentage point more likely to express strong interest (some 
interest) in purchasing EVs, compared to those with limited exposure. These gains in probability comes from a reduction of 
8.5 and 3.9 percentage points from the first two opinion categories. Our 
findings complements \cite{zhao2022determines} and \citet{Scherrer-2023}, who 
demonstrate that greater exposure to EV information through peers or media positively influences attitudes toward EVs, and with 
\citet{Krause-etal-2013}, who underscore the role of accurate 
information in shaping perceptions.

Going forward, we discuss the covariate effects only on the first and last 
categories as they show the largest changes in probability. For the 2021 
survey, the belief that EVs are environmentally 
better than gas-powered vehicles is associated with a 15.5 percentage point 
increase in strong interest and a 19.5 percentage point decrease in no 
interest, underscoring the importance of environmental perceptions in shaping 
purchase intent. This finding complements \citet{Carley-etal-2013}, 
\citet{Krause-etal-2013},  \citet{Scherrer-2023}, and 
\citet{mamkhezri2025public} who show that environmentally sensitive 
respondents tend to have a more favorable opinion of EVs. Here the effect 
reflects perceptions of EVs' benefits over gasoline vehicles rather than 
general pro-environmental attitudes\footnote{Relatedly, 
\citet{Buhmann-Criado-2023} find that reputation-conscious consumers prefer 
EVs only when their prices exceed those of internal combustion engine 
vehicles, suggesting that, for these consumers, reputation takes precedence 
over environmental concerns.}.

%----------------------------  Table 2 ---------------------------------------
\begin{table}[!t]
\centering \setlength{\tabcolsep}{6pt} 
\setlength{\extrarowheight}{1.25pt}
\setlength\arrayrulewidth{1pt}
\caption{{\normalsize{Results from ordinal probit models: posterior mean 
		(Mean) and standard deviation (Std) of the parameters, and logarithm 
		of marginal likelihood ($\ln$ ML), from April 
		2021; May 2022; and May--June 2023} survey data.}}
\begin{tabular}{l rrr rrr rrr }
\toprule
%------------------------------------------------------------------------------
&& \multicolumn{2}{c}{April 2021} && \multicolumn{2}{c}{May 
	2022} && \multicolumn{2}{c}{May--June 2023} \\
%----------------------
\cmidrule{3-4} \cmidrule{6-7}  \cmidrule{9-10} 
%------------------------------------------------------------------------------
&&  Mean & Std &&  Mean & Std &&  Mean & Std \\
\midrule
%--------------------------
\textrm{Intercept}       
&& $ -0.28$  & $0.07$  && $  0.36$ & $0.09$  && $ -0.05$ & $0.09$ \\
%--------------------------
\textrm{EV Info}       
&& $  0.43$  & $0.02$  && $ ..$ & $..$  && $..$ & $..$ \\
%--------------------------
\textrm{Env Better}       
&& $  0.79$  & $0.03$  && $ ..$ & $..$  && $..$ & $..$ \\
%--------------------------
\textrm{EV Infra}       
&& $..$  & $..$  && $..$ & $..$  && $ 0.82$ & $0.04$ \\
%--------------------------
\textrm{GACC (TM)}       
&& $ -0.79$  & $0.04$  && $ -1.13$ & $0.04$  && $-1.07$ & $0.04$ \\
%--------------------------
\textrm{GACC (RA)}       
&& $ -0.52$  & $0.03$  && $ -0.63$ & $0.03$  && $-0.49$ & $0.04$ \\
%--------------------------
\textrm{EV Owner}       
&& $  0.79$  & $0.04$  && $  0.86$ & $0.05$  && $ 0.82$ & $0.04$ \\
%--------------------------
\textrm{Age $< 50$}       
&& $  0.16$  & $0.02$  && $  0.17$ & $0.03$  && $ 0.15$ & $0.03$ \\
%--------------------------
\textrm{Female}       
&& $ -0.02$  & $0.02$  && $ -0.11$ & $0.03$  && $-0.12$ & $0.03$ \\
%--------------------------
\textrm{Grad \& Above}       
&& $  0.38$  & $0.04$  && $  0.45$ & $0.04$  && $ 0.49$ & $0.04$ \\
%--------------------------
\textrm{Some College}       
&& $  0.18$  & $0.04$  && $  0.23$ & $0.04$  && $ 0.21$ & $0.04$ \\
%--------------------------
\textrm{Married}       
&& $ ..$  & $..$       && $  0.06$ & $0.03$  && $ 0.02$ & $0.04$ \\
%--------------------------
\textrm{Metropolitan}       
&& $  0.24$  & $0.03$  && $  0.25$ & $0.04$  && $ 0.20$ & $0.04$ \\
%--------------------------
\textrm{Midwest}       
&& $ -0.08$  & $0.03$  && $ -0.05$ & $0.04$  && $-0.07$ & $0.04$ \\
%--------------------------
\textrm{South}       
&& $ -0.05$  & $0.03$  && $ -0.02$ & $0.04$  && $-0.07$ & $0.04$ \\
%--------------------------
\textrm{West}       
&& $  0.19$  & $0.03$  && $  0.19$ & $0.04$  && $ 0.18$ & $0.04$ \\
%--------------------------
\textrm{US Born}       
&& $ -0.21$  & $0.04$  && $ -0.33$ & $0.04$  && $-0.22$ & $0.04$ \\
%--------------------------
\textrm{Black}       
&& $ -0.28$  & $0.04$  && $ -0.14$ & $0.06$  && $-0.18$ & $0.05$ \\
%--------------------------
\textrm{Other Races}       
&& $  0.05$  & $0.03$  && $  0.01$ & $0.04$  && $-0.03$ & $0.04$ \\
%--------------------------
\textrm{Democrat}       
&& $  0.59$  & $0.03$  && $  0.65$ & $0.04$  && $ 0.75$ & $0.04$ \\
%--------------------------
\textrm{Ind \& Others}       
&& $  0.40$  & $0.03$  && $  0.36$ & $0.03$  && $ 0.43$ & $0.03$ \\
%--------------------------
\textrm{Income/10000 (USD)}       
&& $  0.01$  & $0.01$  && $  0.02$ & $0.01$  && $ 0.03$ & $0.01$ \\
%--------------------------
\textrm{$\delta_{1}$}       
&& $  0.01$  & $0.02$  && $ -0.10$ & $0.02$  && $-0.11$ & $0.02$ \\
%--------------------------
\textrm{$\delta_{2}$}       
&& $  0.12$  & $0.01$  && $ -0.00$ & $0.02$  && $ 0.04$ & $0.02$ \\
\midrule 
\textrm{$\ln$ ML} && \multicolumn{2}{l}{$-12415$} && 
\multicolumn{2}{l}{$-8301$} && \multicolumn{2}{l}{$-7914$} \\
\bottomrule
\end{tabular}
\label{Table:EVProbitResults}
\end{table}
%------------------------------------------------------------------------------

Confidence in charging infrastructure, assessed in the 2023 survey, is one of
the strongest predictors of EV interest. Respondents very or extremely 
confident that the US will build the necessary infrastructure are 19.1 (18.8) 
percentage points more likely to express strong (no) interest in purchasing 
EVs, compared to those who report lower confidence. This finding 
complements research, such as \citet{Carley-etal-2013} and 
\citet{li2017market}, which highlight the importance of existing 
infrastructure. 

Respondents who think the government is doing too much to address climate 
change (GACC, TM) are 14.8–19.7 percentage points less likely to express 
strong interest in purchasing EVs and 19.7–33.4 percentage points more likely 
to report no interest. Those who believe the government is doing the right 
amount (GACC, RA) also express lower interest, though to a lesser extent. 
This relationship has not, to our knowledge, been examined in prior 
work.

Across all surveys, previous EV ownership is consistently linked with 
a higher purchase interest. Respondents who have owned an EV or hybrid are 
19.0–23.1 percentage points more likely to report strong interest in buying 
an EV and 13.2–18.1 percentage points less likely to express no interest than
those without prior ownership. This pattern, also noted by 
\citet{Carley-etal-2013} and \citet{Buhmann-Criado-2023}, suggests that 
firsthand experience reduces uncertainty and fosters positive perceptions of 
EVs. However, selection effects may also be at play, as these 
individuals are likely early adopters with a strong pre-existing interest in 
EVs.

Respondents under age 50 are 2.8–3.7 percentage points more likely to express 
strong interest and 3.3–4.1 percentage points less likely to report no 
interest in purchasing EVs than older respondents. Our finding aligns with 
\citet{zhao2022determines}, \citet{Carley-etal-2013}, 
\citet{Krause-etal-2013}, \citet{Scherrer-2023}, and 
\citet{mamkhezri2025public}. For instance, \citet{zhao2022determines} finds 
that EV buyers in Shanghai are mostly aged 31-40. 

Results from the 2022 and 2023 surveys also show that female respondents are 
are 2.3–2.4 percentage points less likely to express strong interest, and 
2.7–3.0 percentage points more likely to report no interest, than males. This 
finding is consistent with \citet{Carley-etal-2013}, 
\citet{Krause-etal-2013}, \citet{Scherrer-2023}, \citet{Buhmann-Criado-2023}, 
and \citet{mamkhezri2025public}, which also find that men show
stronger preference for EVs.

Higher education is also linked to greater EV interest. Respondents with a 
graduate degree are 8.6–10.0 percentage points more likely to report 
strong interest and 8.1–12.6 percentage points less likely to express no 
interest than those with a high school diploma or less. Similar, 
though smaller, effects are observed for respondents with some college 
education. Such a relationship is 
similarly documented in \citet{Carley-etal-2013}, \citet{Krause-etal-2013}, 
\citet{Buhmann-Criado-2023}, and \citet{mamkhezri2025public}. 

Metropolitan residents are 3.6–5.3 percentage points more likely to report 
strong interest and 5.1–6.6 percentage points less likely to express no 
interest. This finding aligns with \citet{Buhmann-Criado-2023} and 
\citet{mamkhezri2025public}, who also report higher EV preference 
among urban residents. Moreover, respondents in the West are 3.6–4.4 
percentage points more likely to report strong interest than those in 
the Northeast; those in the Midwest and South show no significant 
difference from Northeastern respondents. This pattern likely reflects 
California's leadership in EV adoption, driven in part 
ZEV mandates, the Clean Vehicle Rebate Program, and a robust 
charging network.\footnote{See 
https://www.eia.gov/todayinenergy/detail.php?id=61082}

US-born respondents are 4.4–7.8 percentage points less likely to express 
strong interest and 4.2–7.8 percentage points more likely to express no 
interest than non-US-born respondents. Moreover, Black respondents are 
2.9–5.9 percentage points less likely to express strong interest and 3.5–6.1 
percentage points more likely to express no interest, as compared to White 
respondents. Other non-white respondents 
show similar preferences to White respondents. These differences may reflect 
cultural attitudes, driving patterns, cost perceptions, or structural 
barriers such as limited charging access  \citep{hsu2021public}.

Political affiliation is a consistent and significant predictor of EV 
purchase intent. Relative to Republicans, Democrats are 13.9–15.6 percentage 
points more likely to report strong interest in purchasing an EV and 
12.1–18.8 percentage points less likely to report no interest, while 
Independents and others are about 8 percentage points more likely to express 
strong interest. These findings align with \citet{mamkhezri2025public} and 
likely reflect Democrats' stronger perceptions of EVs as environmentally 
responsible \citep{sintov2020partisan}. They also reflect 
the stark policy differences between the two major political parties, 
particularly during the tenures of Presidents Biden and Trump. During his 
first term (2017–2021), Trump rolled back Obama-era fuel efficiency 
standards, slowing regulatory pressure on automakers to transition toward 
EVs. In his second term, the Trump administration has again taken a 
deregulatory approach. In contrast, the Biden administration adopted an 
aggressive pro-EV stance, prioritizing stricter fuel efficiency and emissions 
standards and allocating \$7.5 billion through the Infrastructure Investment 
and Jobs Act to develop a nationwide network of 500,000 EV charging stations.

%----------------------------  Table 3 ---------------------------------------
\begin{landscape}
\begin{table}[!h]
\centering \small \setlength{\tabcolsep}{3pt} 
\setlength{\extrarowheight}{3pt}
\setlength\arrayrulewidth{1pt}
\caption{{\normalsize{Covariate effects (CE) from ordinal probit models. CE 
marked with $\times$ indicate parameters that are not statistically different 
from zero at 95\% probability level.}}}
\begin{tabular}{l rrrr rrrrr rrrrr rrrr}
\toprule
&& \multicolumn{4}{c}{April 2021} 
&& \multicolumn{4}{c}{May 2022} 
&& \multicolumn{4}{c}{May--June 2023}  \\
\cmidrule{3-6} \cmidrule{8-11} \cmidrule{13-16}
%------------------------------------------------------------------------------
%&& \multicolumn{2}{c}{$\Delta P(y=1)$} && \multicolumn{2}{c}{0th}}
%&& \multicolumn{2}{c}{\textsc{75th}} &&& \multicolumn{2}{c}{\textsc{25th}}
%&& \multicolumn{2}{c}{\textsc{50th}} && \multicolumn{2}{c}{\textsc{75th}}  
%&&& \multicolumn{2}{c}{\textsc{25th}}
%&& \multicolumn{2}{c}{\textsc{50th}} && \multicolumn{2}{c}{\textsc{75th}} \\
%\cmidrule{3-4} \cmidrule{6-7}  \cmidrule{9-10} \cmidrule{13-14}   
%\cmidrule{16-17} \cmidrule{19-20} \cmidrule{23-24} \cmidrule{26-27} 
%\cmidrule{29-30} 
%------------------------------------------------------------------------------
&&  $\Delta$P(y=1)  & $\Delta$P(y=2)  & $\Delta$P(y=3)  & $\Delta$P(y=4)
&&  $\Delta$P(y=1)  & $\Delta$P(y=2)  & $\Delta$P(y=3)  & $\Delta$P(y=4)
&&  $\Delta$P(y=1)  & $\Delta$P(y=2)  & $\Delta$P(y=3)  & $\Delta$P(y=4) \\
\midrule
%--------------------------
\textrm{EV Info}       
&& $ -0.0851$  & $-0.0388$ & $0.0217$   & $0.1022$   
&& $ ..$       & $ .. $    & $ .. $     & $..$   
&& $ ..$       & $ .. $    & $ .. $     & $..$     \\
%--------------------------
\textrm{Env Better}       
&& $ -0.1955$  & $-0.0565$ & $ 0.0970$  & $0.1551$   
&& $ ..$       & $ .. $    & $ .. $     & $..$    
&& $ ..$       & $ .. $    & $ .. $     & $..$     \\
%--------------------------
\textrm{EV Infra}       
&& $ ..$       & $ .. $    & $ .. $     & $..$     
&& $ ..$       & $ .. $    & $ .. $     & $..$      
&& $ -0.1880$  & $-0.0553$ & $ 0.0526$  & $0.1908$  \\
%--------------------------
\textrm{GACC (TM)}       
&& $  0.1971$  & $0.0453$  & $ -0.0945$ & $-0.1478$   
&& $  0.3341$  & $0.0133$  & $ -0.1507$ & $-0.1968$   
&& $  0.3170$  & $-0.0153$ & $ -0.1395$ & $-0.1622$  \\
%--------------------------
\textrm{GACC (RA)}       
&& $  0.1180$  & $0.0391$  & $ -0.0490$ & $-0.1082$   
&& $  0.1649$  & $0.0267$  & $ -0.0639$ & $-0.1278$   
&& $  0.1282$  & $0.0079$  & $ -0.0486$ & $-0.0875$  \\
%--------------------------
\textrm{EV Owner}       
&& $ -0.1321$  & $-0.0849$ & $  0.0051$ & $0.2119$   
&& $ -0.1749$  & $-0.0794$ & $  0.0231$ & $0.2312$   
&& $ -0.1814$  & $-0.0513$ & $  0.0421$ & $0.1905$  \\
%--------------------------
\textrm{Age $< 50$}       
&& $ -0.0326$  & $-0.0130$ & $  0.0099$ & $0.0356$   
&& $ -0.0410$  & $-0.0090$ & $  0.0129$ & $0.0371$   
&& $ -0.0371$  & $-0.0036$  & $ 0.0126$ & $0.0282$  \\
%--------------------------
\textrm{Female}       
&& $  \times$  & $\times$  & $\times  $ & $\times$   
&& $  0.0268$  & $0.0055$  & $ -0.0082$ & $-0.0240$   
&& $  0.0301$  & $0.0027$  & $ -0.0100$ & $-0.0228$  \\
%--------------------------
\textrm{Grad \& Above}       
&& $ -0.0815$  & $-0.0322$ & $  0.0276$ & $0.0860$   
&& $ -0.1139$  & $-0.0240$ & $  0.0378$ & $0.1001$   
&& $ -0.1264$  & $-0.0130$ & $ 0.0445$ & $0.0949$  \\
%--------------------------
\textrm{Some College}       
&& $ -0.0373$  & $-0.0152$ & $  0.0099$ & $0.0426$   
&& $ -0.0541$  & $-0.0118$ & $  0.0156$ & $0.0503$   
&& $ -0.0517$  & $-0.0052$ & $ -0.0166$ & $0.0403$  \\
%--------------------------
\textrm{Married}       
&& $ ..$       & $ .. $    & $ .. $     & $..$     
&& $ -0.0154$  & $-0.0029$ & $  0.0050$ & $0.0134$   
&& $  \times$  & $\times$  & $\times  $ & $\times$   \\
%--------------------------
\textrm{Metropolitan}       
&& $ -0.0534$  & $-0.0179$ & $  0.0191$ & $0.0522$   
&& $ -0.0662$  & $-0.0099$ & $  0.0232$ & $0.0528$   
&& $ -0.0512$  & $-0.0027$ & $  0.0183$ & $0.0356$  \\
%--------------------------
\textrm{Midwest}       
&& $  0.0162$  & $0.0061$  & $ -0.0051$ & $-0.0172$   
&& $  \times$  & $\times$  & $\times  $ & $\times$   
&& $  \times$  & $\times$  & $\times  $ & $\times$   \\
%--------------------------
\textrm{South}       
&& $  \times$  & $\times$  & $\times  $ & $\times$   
&& $  \times$  & $\times$  & $\times  $ & $\times$   
&& $  \times$  & $\times$  & $\times  $ & $\times$   \\
%--------------------------
\textrm{West}       
&& $ -0.0378$  & $-0.0164$ & $  0.0101$ & $0.0441$   
&& $ -0.0457$  & $-0.0109$ & $  0.0135$ & $0.0431$   
&& $ -0.0454$  & $-0.0053$ & $  0.0148$ & $0.0360$  \\
%--------------------------
\textrm{US Born}       
&& $  0.0416$  & $0.0186$  & $ -0.0106$ & $-0.0497$   
&& $  0.0777$  & $0.0217$  & $ -0.0214$ & $-0.0781$   
&& $  0.0551$  & $0.0071$  & $ -0.0178$ & $-0.0444$  \\
%--------------------------
\textrm{Black}       
&& $  0.0614$  & $0.0200$  & $ -0.0220$ & $-0.0594$   
&& $  0.0349$  & $0.0058$  & $ -0.0116$ & $-0.0291$   
&& $  0.0460$  & $0.0026$  & $ -0.0160$ & $-0.0326$  \\
%--------------------------
\textrm{Other Races}       
&& $  \times$  & $\times$  & $\times  $ & $\times$   
&& $  \times$  & $\times$  & $\times  $ & $\times$   
&& $  \times$  & $\times$  & $\times  $ & $\times$   \\
%--------------------------
\textrm{Democrat}       
&& $ -0.1211$  & $-0.0578$ & $  0.0403$ & $0.1386$   
&& $ -0.1569$  & $-0.0497$ & $  0.0513$ & $0.1553$   
&& $ -0.1882$  & $-0.0353$ & $  0.0677$ & $0.1558$  \\
%--------------------------
\textrm{Ind \& Others}       
&& $ -0.0808$  & $-0.0310$ & $  0.0219$ & $0.0898$   
&& $ -0.0890$  & $-0.0187$ & $  0.0266$ & $0.0811$   
&& $ -0.1091$  & $-0.0093$ & $  0.0356$ & $0.0827$  \\
%--------------------------
\textrm{$\Delta$I$=\$50,000$}       
&& $  \times$  & $\times$  & $\times  $ & $\times$   
&& $ -0.0223$  & $-0.0051$ & $  0.0066$ & $0.0209$   
&& $ -0.0363$  & $-0.0045$ & $  0.0116$ & $0.0291$  \\
%------------------------------------------------------------------------------
\bottomrule
\end{tabular}
\label{Table:OPCE}
\end{table}
\end{landscape}
%------------------------------------------------------------------------------

Income has a positive effect on EV purchase intent. Conditional on other 
factors, an additional \$50,000 in income changes the probability of 
strong interest (no interest) by 2.1–2.9 (2.2–3.6) percentage 
points. Our results find support in \citet{zhao2022determines} and 
\citet{mamkhezri2025public}, who report greater support for EVs among 
higher-income individuals. The small covariate effect may reflect 
the growing availability of EVs across various market segments and the 
subsidies for low-income consumers, which reduce affordability 
barriers compared to earlier market years. 

Overall, the results show that interest in EVs varies by environmental attitudes, experience with EVs, confidence in charging infrastructure, and demographics. Some patterns likely reflect personal preferences; others may reflect barriers such as limited awareness or charging access. 

%-----------------------------------------------------------------------------
\subsection{Results from Ordinal Quantile Models}

We now present the results from ordinal quantile models \citep{Rahman-2016} 
that shift attention from average probabilities of outcomes to those at specific quantiles, conditional on the 
covariates. To our knowledge, this aspect has not been examined in prior work 
on EVs. Table~\ref{Table:EVQuantResults} presents 
the posterior means and standard deviations of the parameters from the 
ordinal quantile models estimated separately for each survey; along with the 
logarithm of marginal likelihood which can be used for 
Bayesian model comparison \citep{Maheshwari-Rahman-2023}. Note that the 
coefficients in Table~\ref{Table:EVQuantResults} are not directly comparable 
to those in Table~\ref{Table:EVProbitResults} for two main reasons: (a) the 
models operate on different scales due to differences in the variance of the 
error terms, and (b) they address distinct inferential goals$-$the ordinal 
probit model focuses on the average probability of outcomes, whereas the 
ordinal quantile model targets the probability of outcomes at specific 
quantiles.

Table~\ref{Table:QuantCE} reports 
the corresponding covariate effects for all variables whose parameters are 
statistically different from zero at 95\% probability level. We see that 
covariate effects at the 50th percentile align closely with those from the 
ordinal probit model, but across all surveys, those at the 20th and 80th 
quantiles are noticeably different. This provides compelling evidence that 
the impact of covariates on purchase intent is not uniform but varies across 
the distribution of purchase intent, underscoring the value of quantile 
modeling in capturing such heterogeneity. Next, we emphasize the key 
differences across quantiles.

Although having heard or read extensively about EVs corresponds to higher 
purchase intent, the relationship is stronger among respondents more inclined 
toward adoption. In the 2021 survey and at the 80th percentile of purchase 
intent, respondents who have heard or read extensively about EVs are 12.6 
percentage points more likely to express strong intent than those who have read or heard little to nothing. At the 50th and 20th 
percentiles of purchase intent, the difference is 9.9 and 6.1 percentage 
points, respectively.

\begin{landscape}
%----------------------------  Table 4 ---------------------------------------
\begin{table}[!t]
\centering \small \setlength{\tabcolsep}{2.5pt} 
\setlength{\extrarowheight}{2.5pt}
\setlength\arrayrulewidth{1pt}
\caption{{\normalsize{Results from quantile ordinal model: posterior mean 
(Mean) and standard deviation (Std) of the parameters, along with the 
logarithm of marginal likelihood ($\ln$ ML), from 
April 2021, May 2022, and May--Jun 2023 survey data.}}}
\begin{tabular}{l rrr rrr rrr rrr rrrr rr rrr rrrr rrrrr }
\toprule
& & \multicolumn{8}{c}{April 2021} &&& 
\multicolumn{8}{c}{May 2022} &&& \multicolumn{8}{c}{May--June 2023}  
\\
\cmidrule{3-10}  \cmidrule{13-20} \cmidrule{23-30}
%------------------------------------------------------------------------------
&& \multicolumn{2}{c}{\textsc{20th}} && \multicolumn{2}{c}{\textsc{50th}}
&& \multicolumn{2}{c}{\textsc{80th}} &&& \multicolumn{2}{c}{\textsc{20th}}
&& \multicolumn{2}{c}{\textsc{50th}} && \multicolumn{2}{c}{\textsc{80th}}  
&&& \multicolumn{2}{c}{\textsc{20th}}
&& \multicolumn{2}{c}{\textsc{50th}} && \multicolumn{2}{c}{\textsc{80th}} \\
\cmidrule{3-4} \cmidrule{6-7}  \cmidrule{9-10} \cmidrule{13-14}   
\cmidrule{16-17} \cmidrule{19-20} \cmidrule{23-24} \cmidrule{26-27} 
\cmidrule{29-30} 
%------------------------------------------------------------------------------
&&  Mean & Std && Mean & Std  &&  Mean & Std 
&&& Mean & Std && Mean & Std  &&  Mean & Std  
&&& Mean & Std && Mean & Std  &&  Mean & Std  &    \\
\midrule
%--------------------------
\textrm{Intercept}       
&&  $ -3.60$  & $0.25$  && $ -0.58$ & $0.17$  && $  2.60$ & $0.25$ 
&&& $ -2.15$  & $0.30$  && $  0.80$ & $0.21$  && $  4.64$ & $0.29$  
&&& $ -2.81$  & $0.30$  && $ -0.05$ & $0.20$  && $  2.77$ & $0.28$ \\
%--------------------------
\textrm{EV Info}       
&&  $  1.28$  & $0.08$  && $  0.98$ & $0.06$  && $  1.64$ & $0.08$ 
&&& $  ..$    & $..$    && $  ..$   & $..$    && $ ..$    & $..$ 
&&& $  ..$    & $..$    && $  ..$   & $..$    && $ ..$    & $..$ \\
%--------------------------
\textrm{Env Better}       
&&  $  2.52$  & $0.11$  && $  1.80$ & $0.07$  && $  2.85$ & $0.10$ 
&&& $  ..$    & $..$    && $  ..$   & $..$    && $ ..$    & $..$ 
&&& $  ..$    & $..$    && $  ..$   & $..$    && $ ..$    & $..$ \\
%--------------------------
\textrm{EV Infra}       
&&  $  ..$    & $..$    && $  ..$   & $..$    && $ ..$    & $..$ 
&&& $  ..$    & $..$    && $  ..$   & $..$    && $ ..$    & $..$ 
&&& $ 2.83$   & $0.13$  && $  1.92$ & $0.10$  && $ 2.71$  & $0.14$ \\
%--------------------------
\textrm{GACC (TM)}       
&&  $ -2.81$  & $0.13$  && $ -1.96$ & $0.09$  && $ -2.49$ & $0.13$ 
&&& $ -3.79$  & $0.16$  && $ -2.63$ & $0.10$  && $ -3.70$ & $0.14$  
&&& $ -4.00$  & $0.17$  && $ -2.41$ & $0.10$  && $ -3.45$ & $0.14$ \\
%--------------------------
\textrm{GACC (RA)}       
&&  $ -1.79$  & $0.10$  && $ -1.24$ & $0.07$  && $ -1.55$ & $0.11$ 
&&& $ -1.94$  & $0.12$  && $ -1.48$ & $0.08$  && $ -2.03$ & $0.12$  
&&& $ -1.77$  & $0.12$  && $ -1.15$ & $0.08$  && $ -1.48$ & $0.12$ \\
%--------------------------
\textrm{EV Owner}       
&&  $  2.80$  & $0.13$  && $  1.76$ & $0.10$  && $  2.91$ & $0.16$ 
&&& $  2.94$  & $0.16$  && $  1.83$ & $0.11$  && $  3.31$ & $0.20$  
&&& $  2.65$  & $0.16$  && $  1.99$ & $0.11$  && $  3.08$ & $0.19$ \\
%--------------------------
\textrm{Age$< 50$}       
&&  $  0.40$  & $0.08$  && $  0.36$ & $0.05$  && $  0.58$ & $0.08$ 
&&& $  0.50$  & $0.09$  && $  0.37$ & $0.07$  && $  0.63$ & $0.10$  
&&& $  0.42$  & $0.11$  && $  0.34$ & $0.07$  && $  0.55$ & $0.10$ \\
%--------------------------
\textrm{Female}       
&&  $ -0.11$  & $0.08$  && $ -0.02$ & $0.05$  && $ -0.03$ & $0.07$ 
&&& $ -0.36$  & $0.10$  && $ -0.22$ & $0.06$  && $ -0.33$ & $0.09$  
&&& $ -0.49$  & $0.10$  && $ -0.31$ & $0.07$  && $ -0.27$ & $0.09$ \\
%--------------------------
\textrm{Grad \& Above}       
&&  $  1.42$  & $0.14$  && $ -0.93$ & $0.08$  && $  1.17$ & $0.13$ 
&&& $  1.93$  & $0.16$  && $  1.04$ & $0.10$  && $  1.19$ & $0.13$  
&&& $  1.94$  & $0.16$  && $  1.09$ & $0.10$  && $  1.43$ & $0.13$ \\
%--------------------------
\textrm{Some College}       
&&  $  0.66$  & $0.14$  && $  0.47$ & $0.09$  && $  0.57$ & $0.12$ 
&&& $  1.08$  & $0.16$  && $  0.53$ & $0.10$  && $  0.49$ & $0.13$  
&&& $  0.92$  & $0.16$  && $  0.42$ & $0.10$  && $  0.53$ & $0.13$ \\
%--------------------------
\textrm{Married}       
&&  $ ..$     & $..$    && $ ..$    & $..$    && $ ..$    & $ ..$ 
&&& $  0.22$  & $0.13$  && $  0.12$ & $0.08$  && $  0.19$ & $0.12$  
&&& $  0.02$  & $0.13$  && $  0.08$ & $0.08$  && $  0.12$ & $0.11$ \\
%--------------------------
\textrm{Metropolitan}       
&&  $  0.77$  & $0.12$  && $  0.64$ & $0.08$  && $  0.71$ & $0.12$ 
&&& $  0.91$  & $0.16$  && $  0.59$ & $0.10$  && $  0.79$ & $0.13$  
&&& $  0.79$  & $0.16$  && $  0.46$ & $0.10$  && $  0.61$ & $0.13$ \\
%--------------------------
\textrm{Midwest}       
&&  $ -0.31$  & $0.12$  && $ -0.19$ & $0.08$  && $ -0.24$ & $0.11$ 
&&& $ -0.09$  & $0.15$  && $ -0.09$ & $0.10$  && $ -0.11$ & $0.14$  
&&& $ -0.18$  & $0.16$  && $ -0.15$ & $0.10$  && $ -0.21$ & $0.14$ \\
%--------------------------
\textrm{South}       
&& $  -0.21$  & $0.10$  && $ -0.12$ & $0.07$  && $ -0.17$ & $0.10$ 
&&& $  0.00$  & $0.13$  && $ -0.02$ & $0.09$  && $ -0.06$ & $0.13$  
&&& $ -0.22$  & $0.15$  && $ -0.22$ & $0.09$  && $ -0.17$ & $0.13$ \\
%--------------------------
\textrm{West}       
&&  $  0.57$  & $0.12$  && $  0.47$ & $0.08$  && $  0.70$ & $0.12$ 
&&& $  0.46$  & $0.15$  && $  0.45$ & $0.10$  && $  0.89$ & $0.14$  
&&& $  0.63$  & $0.16$  && $  0.46$ & $0.10$  && $  0.66$ & $0.15$ \\
%--------------------------
\textrm{US Born}       
&&  $ -0.63$  & $0.12$  && $ -0.49$ & $0.09$  && $ -0.80$ & $0.14$ 
&&& $ -0.89$  & $0.15$  && $ -0.67$ & $0.10$  && $ -1.25$ & $0.16$  
&&& $ -0.91$  & $0.16$  && $ -0.45$ & $0.11$  && $ -0.67$ & $0.15$ \\
%--------------------------
\textrm{Black}       
&&  $ -1.23$  & $0.15$  && $ -0.70$ & $0.10$  && $ -0.56$ & $0.14$ 
&&& $ -0.86$  & $0.20$  && $ -0.25$ & $0.14$  && $ -0.06$ & $0.18$  
&&& $ -0.94$  & $0.16$  && $ -0.42$ & $0.11$  && $ -0.36$ & $0.16$ \\
%--------------------------
\textrm{Other Races}       
&&  $  0.08$  & $0.10$  && $  0.14$ & $0.07$  && $  0.25$ & $0.11$ 
&&& $ -0.19$  & $0.14$  && $  0.07$ & $0.10$  && $  0.22$ & $0.13$  
&&& $ -0.30$  & $0.15$  && $  0.00$ & $0.10$  && $  0.02$ & $0.14$ \\
%--------------------------
\textrm{Democrat}       
&&  $  2.00$  & $0.11$  && $  1.33$ & $0.08$  && $  1.94$ & $0.12$ 
&&& $  2.30$  & $0.14$  && $  1.45$ & $0.09$  && $  1.90$ & $0.13$  
&&& $  2.49$  & $0.14$  && $  1.71$ & $0.10$  && $  2.40$ & $0.14$ \\
%--------------------------
\textrm{Ind \& Others}       
&&  $  1.29$  & $0.10$  && $  0.88$ & $0.07$  && $  1.42$ & $0.11$ 
&&& $  1.32$  & $0.12$  && $  0.80$ & $0.08$  && $  1.06$ & $0.11$  
&&& $  1.40$  & $0.13$  && $  0.95$ & $0.08$  && $  1.44$ & $0.12$ \\
%--------------------------
\textrm{Income/10000 (USD)}       
&&  $  0.04$  & $0.01$  && $  0.02$ & $0.01$  && $  0.00$ & $0.01$ 
&&& $  0.07$  & $0.02$  && $  0.05$ & $0.01$  && $  0.05$ & $0.02$  
&&& $  0.12$  & $0.02$  && $  0.06$ & $0.01$  && $  0.08$ & $0.02$ \\
%--------------------------
\textrm{$\delta_{1}$}       
&&  $  1.16$  & $0.02$  && $  0.91$ & $0.02$  && $  1.45$ & $0.02$ 
&&& $  1.09$  & $0.02$  && $  0.76$ & $0.02$  && $  1.25$ & $0.02$  
&&& $  1.14$  & $0.02$  && $  0.74$ & $0.02$  && $  1.21$ & $0.02$ \\
%--------------------------
\textrm{$\delta_{2}$}       
&&  $  1.53$  & $0.02$  && $  0.95$ & $0.02$  && $  1.35$ & $0.02$ 
&&& $  1.46$  & $0.02$  && $  0.84$ & $0.02$  && $  1.16$ & $0.02$  
&&& $  1.52$  & $0.02$  && $  0.93$ & $0.02$  && $  1.19$ & $0.02$ \\
\midrule
%--------------------------
\textrm{$\ln$ ML}       
&&  \multicolumn{2}{l}{$-12871$}  && \multicolumn{2}{l}{$-12529$}  && 
\multicolumn{2}{l}{$-12901$}
&&& \multicolumn{2}{l}{$-8563$}  && \multicolumn{2}{l}{$-8357$}  && 
\multicolumn{2}{l}{$-8624$} 
&&& \multicolumn{2}{l}{$-8237$}  && \multicolumn{2}{l}{$-7981$}  && 
\multicolumn{2}{l}{$-8194$} \\
%------------------------------------------------------------------------------
\bottomrule
\end{tabular}
\label{Table:EVQuantResults}
\end{table}
%------------------------------------------------------------------------------
\end{landscape}
%------------------------------------------------------------------------------

%----------------------------  Table 5 ---------------------------------------
\begin{landscape}
\begin{table}[!h]
\centering \footnotesize \setlength{\tabcolsep}{3pt} 
\setlength{\extrarowheight}{1.5pt}
\setlength\arrayrulewidth{1pt}
\caption{{\normalsize{Covariate effects (CE) from quantile analysis. CE 
marked with $\times$ indicate parameters that are not statistically different 
from zero at 95\% probability level.}}}
\begin{tabular}{l rrrr rrrrr rrrrr rrrr}
\toprule
&& \multicolumn{12}{c}{April 2021 Survey}  \\
\cmidrule{3-16}
&& \multicolumn{4}{c}{\textsc{20th Quantile}} 
&& \multicolumn{4}{c}{\textsc{50th Quantile}} 
&& \multicolumn{4}{c}{\textsc{80th Quantile}}  \\
\cmidrule{3-6} \cmidrule{8-11} \cmidrule{13-16}
%------------------------------------------------------------------------------
%&& \multicolumn{2}{c}{$\Delta P(y=1)$} && \multicolumn{2}{c}{0th}}
%&& \multicolumn{2}{c}{\textsc{75th}} &&& \multicolumn{2}{c}{\textsc{25th}}
%&& \multicolumn{2}{c}{\textsc{50th}} && \multicolumn{2}{c}{\textsc{75th}}  
%&&& \multicolumn{2}{c}{\textsc{25th}}
%&& \multicolumn{2}{c}{\textsc{50th}} && \multicolumn{2}{c}{\textsc{75th}} \\
%\cmidrule{3-4} \cmidrule{6-7}  \cmidrule{9-10} \cmidrule{13-14}   
%\cmidrule{16-17} \cmidrule{19-20} \cmidrule{23-24} \cmidrule{26-27} 
%\cmidrule{29-30} 
%------------------------------------------------------------------------------
&&  $\Delta$P(y=1)  & $\Delta$P(y=2)  & $\Delta$P(y=3)  & $\Delta$P(y=4)
&&  $\Delta$P(y=1)  & $\Delta$P(y=2)  & $\Delta$P(y=3)  & $\Delta$P(y=4)
&&  $\Delta$P(y=1)  & $\Delta$P(y=2)  & $\Delta$P(y=3)  & $\Delta$P(y=4) \\
\midrule
%--------------------------
\textrm{EV Info}       
&& $ -0.0722$  & $-0.0421$  & $ 0.0527$ & $0.0615$   
&& $ -0.0792$  & $-0.0374$  & $ 0.0172$ & $0.0995$   
&& $ -0.0722$  & $-0.0453$  & $-0.0089$ & $0.1265$   \\
%--------------------------
\textrm{Env Better}       
&& $ -0.1958$  & $-0.0191$  & $ 0.1135$ & $0.1014$   
&& $ -0.1771$  & $-0.0730$  & $ 0.1122$ & $0.1379$   
&& $ -0.1384$  & $-0.1127$  & $ 0.0789$ & $0.1722$   \\
%--------------------------
\textrm{GACC (TM)}       
&& $  0.2192$  & $0.0097$  & $ -0.1217$ & $-0.1072$   
&& $  0.1974$  & $0.0635$  & $ -0.1203$ & $-0.1406$   
&& $  0.1283$  & $0.0779$  & $ -0.0575$ & $-0.1487$  \\
%--------------------------
\textrm{GACC (RA)}       
&& $  0.1216$  & $0.0312$  & $ -0.0780$ & $-0.0747$   
&& $  0.1097$  & $0.0534$  & $ -0.0613$ & $-0.1018$   
&& $  0.0763$  & $0.0421$  & $ -0.0124$ & $-0.1059$  \\
%--------------------------
\textrm{EV Owner}       
&& $ -0.1357$  & $-0.0896$  & $ 0.0633$ & $0.1620$   
&& $ -0.1212$  & $-0.0757$  & $-0.0113$ & $0.2083$   
&& $ -0.1053$  & $-0.0953$  & $-0.0265$ & $0.2270$  \\
%--------------------------
\textrm{Age $< 50$}       
&& $ -0.0237$  & $-0.0106$  & $ 0.0156$ & $0.0187$   
&& $ -0.0303$  & $-0.0134$  & $ 0.0095$ & $0.0343$   
&& $ -0.0265$  & $-0.0164$  & $ 0.0003$ & $0.0427$  \\
%--------------------------
\textrm{Metropolitan}       
&& $ -0.0483$  & $-0.0174$  & $  0.0318$ & $0.0338$   
&& $ -0.0565$  & $-0.0225$  & $  0.0237$ & $0.0553$   
&& $ -0.0345$  & $-0.0185$  & $  0.0030$ & $0.0500$  \\
%--------------------------
\textrm{Midwest}       
&& $  0.0185$  & $ 0.0080$  & $ -0.0123$ & $-0.0141$   
&& $  0.0161$  & $ 0.0068$  & $ -0.0053$ & $-0.0177$   
&& $  0.0114$  & $ 0.0066$  & $ -0.0003$ & $-0.0177$  \\
%--------------------------
\textrm{South}       
&& $  0.0124$  & $0.0056$  & $ -0.0083$ & $-0.0098$   
&& $  \times$  & $\times$  & $  \times$ & $\times$   
&& $  \times$  & $\times$  & $  \times$ & $\times$   \\
%--------------------------
\textrm{West}       
&& $ -0.0327$  & $-0.0168$  & $ 0.0220$ & $0.0275$   
&& $ -0.0382$  & $-0.0176$  & $ 0.0092$ & $0.0467$   
&& $ -0.0311$  & $-0.0200$  & $-0.0015$ & $0.0526$  \\
%--------------------------
\textrm{Black}       
&& $  0.0784$  & $ 0.0256$  & $ -0.0519$ & $-0.0520$   
&& $  0.0613$  & $ 0.0244$  & $ -0.0259$ & $-0.0598$   
&& $  0.0269$  & $ 0.0141$  & $ -0.0014$ & $-0.0396$  \\
%--------------------------
\textrm{Democrat}       
&& $ -0.1242$  & $-0.0523$  & $ 0.0808$ & $0.0958$   
&& $ -0.1081$  & $-0.0631$  & $ 0.0431$ & $0.1280$   
&& $ -0.0828$  & $-0.0676$  & $ 0.0070$ & $0.1434$  \\
%--------------------------
%------------------------------------------------------------------------------
\midrule
&& \multicolumn{12}{c}{May 2022 Survey}  \\
\cmidrule{3-16}
%--------------------------
\textrm{GACC (TM)}       
&& $  0.3425$  & $-0.0568$  & $ -0.1515$ & $-0.1341$   
&& $  0.3304$  & $ 0.0226$  & $ -0.1778$ & $-0.1752$   
&& $  0.2471$  & $ 0.0963$  & $ -0.1459$ & $-0.1975$  \\
%--------------------------
\textrm{GACC (RA)}       
&& $  0.1541$  & $0.0026$  & $ -0.0790$ & $-0.0777$   
&& $  0.1527$  & $0.0497$  & $ -0.0852$ & $-0.1171$   
&& $  0.1269$  & $0.0392$  & $ -0.0308$ & $-0.1352$  \\
%--------------------------
\textrm{EV Owner}       
&& $ -0.1782$  & $-0.0723$  & $ 0.0880$ & $0.1624$   
&& $ -0.1573$  & $-0.0695$  & $ 0.0121$ & $0.2147$   
&& $ -0.1465$  & $-0.1110$  & $-0.0071$ & $0.2646$  \\
%--------------------------
\textrm{Age $< 50$}       
&& $ -0.0362$  & $-0.0066$  & $ 0.0199$ & $0.0229$   
&& $ -0.0390$  & $-0.0085$  & $ 0.0130$ & $0.0344$   
&& $ -0.0358$  & $-0.0167$  & $ 0.0072$ & $0.0453$  \\
%--------------------------
\textrm{Metropolitan}       
&& $ -0.0684$  & $-0.0054$  & $ 0.0360$ & $0.0378$   
&& $ -0.0643$  & $-0.0109$  & $ 0.0258$ & $0.0493$   
&& $ -0.0483$  & $-0.0165$  & $ 0.0114$ & $0.0535$  \\
%%--------------------------
\textrm{Black}       
&& $  0.0642$  & $0.0060$  & $ -0.0342$ & $-0.0359$   
&& $  \times$  & $\times$  & $ \times$ & $\times$   
&& $  \times$  & $\times$  & $ \times$ & $\times$   \\
%--------------------------
\textrm{Democrat}       
&& $ -0.1719$  & $-0.0295$  & $ 0.0927$ & $0.1087$   
&& $ -0.1437$  & $-0.0574$  & $ 0.0601$ & $0.1410$   
&& $ -0.0996$  & $-0.0659$  & $ 0.0217$ & $0.1438$  \\
%--------------------------
%------------------------------------------------------------------------------
%------------------------------------------------------------------------------
\midrule
&& \multicolumn{12}{c}{May--June 2023 Survey}  \\
\cmidrule{3-16}
%--------------------------
\textrm{EV Infra}       
&& $ -0.1863$  & $-0.0428$  & $ 0.0956$ & $0.1335$   
&& $ -0.1827$  & $-0.0603$  & $ 0.0616$ & $0.1813$   
&& $ -0.1452$  & $-0.0730$  & $ 0.0187$ & $0.1995$  \\
%--------------------------
\textrm{GACC (TM)}       
&& $  0.3323$  & $-0.0635$  & $ -0.1405$ & $-0.1283$   
&& $  0.3103$  & $-0.0250$  & $ -0.1426$ & $-0.1427$   
&& $  0.2512$  & $ 0.0486$  & $ -0.1423$ & $-0.1575$  \\
%--------------------------
\textrm{GACC (RA)}       
&& $  0.1318$  & $-0.0065$  & $ -0.0592$ & $-0.0661$   
&& $  0.1238$  & $ 0.0132$  & $ -0.0570$ & $-0.0799$   
&& $  0.0975$  & $ 0.0183$  & $ -0.0329$ & $-0.0830$  \\
%--------------------------
\textrm{EV Owner}       
&& $ -0.1684$  & $-0.0350$  & $ 0.0727$ & $0.1307$   
&& $ -0.1842$  & $-0.0542$  & $ 0.0459$ & $0.1925$   
&& $ -0.1580$  & $-0.0817$  & $ 0.0183$ & $0.2215$  \\
%--------------------------
\textrm{Age $< 50$}       
&& $ -0.0296$  & $-0.0007$  & $ 0.0131$ & $0.0173$   
&& $ -0.0363$  & $-0.0034$  & $ 0.0139$ & $0.0258$   
&& $ -0.0349$  & $-0.0082$  & $ 0.0102$ & $0.0329$  \\
%--------------------------
\textrm{Metropolitan}       
&& $ -0.0568$  & $ 0.0013$  & $  0.0250$ & $0.0304$   
&& $ -0.0511$  & $-0.0026$  & $  0.0204$ & $0.0333$   
&& $ -0.0408$  & $-0.0067$  & $  0.0129$ & $0.0345$  \\
%--------------------------
\textrm{Black}       
&& $  0.0665$  & $-0.0010$  & $ -0.0296$ & $-0.0359$   
&& $  0.0462$  & $ 0.0029$  & $ -0.0182$ & $-0.0309$   
&& $  0.0239$  & $ 0.0039$  & $ -0.0069$ & $-0.0209$  \\
%--------------------------
\textrm{Democrat}       
&& $ -0.1832$  & $-0.0084$  & $ 0.0851$ & $0.1066$   
&& $ -0.1815$  & $-0.0384$  & $ 0.0798$ & $0.1401$ 
&& $ -0.1393$  & $-0.0660$  & $ 0.0491$ & $0.1562$  \\
%------------------------------------------------------------------------------
\bottomrule
\end{tabular}
\label{Table:QuantCE}
\end{table}
\end{landscape}
%------------------------------------------------------------------------------

Similarly, the relationship between confidence about building charging 
infrastructure (measured in the 2023 survey) and  
purchase intent varies across the distribution of the latter. 
At the 80th percentile, confident respondents are 
20.0 percentage points more likely to express strong purchase intent
than those who are not confident. The 
corresponding differences are 18.1 percentage points at the median and 13.3 
percentage points at the 20th percentile. Conversely, at the 80th percentile, 
those who are confident 
are 14.5 percentage points less likely to express no interest, with 
differences of 18.3 and 18.6 percentage 
points at the median and 20th percentile, respectively. 

Beliefs about EVs' environmental benefits relative to gas-powered vehicles 
(measured in the 2021 survey) also show varying relationships with purchase 
intent across quantiles. Among respondents at the 80th percentile, 
those who believe EVs are environmentally better than gas-powered vehicles 
are 17.2 percentage points more likely to express strong purchase intent than 
those at the same percentile who do not share this belief. The corresponding 
differences are 13.8 percentage points at the median and 10.1 percentage 
points at the 20th percentile. Conversely, such respondents are 13.8 
percentage points less likely to express no interest at the 80th percentile, 
compared to 17.7 and 19.6 percentage points less likely at the median and 
20th percentile, respectively. 

Regarding views on government action, respondents who believe the 
government is doing too much or enough to reduce climate change are 
significantly less likely to express strong purchase interest, 
with this becoming more pronounced at higher percentiles. In the 2021 survey, 
those at the 80th percentile of  purchase intent who hold this belief are 
14.9 percentage points less likely to report strong purchase intent than 
those at the same percentile who do not. The gap is 14.1 and 10.7 percentage 
points at the median and 20th percentile, respectively. Likewise, at 
the 80th percentile, these respondents are 12.8 percentage points more likely 
to express no interest in EVs, with the gap widening to 19.7 and 21.9 
percentage points at the median and 20th percentile, respectively. More 
recent surveys indicate even stronger covariate effects, with those who 
believe the government is doing too much approximately 33 and 25 percentage 
points more likely to express no interest in EVs at the 20th percentile and 
80th percentile, respectively. 

The influence of prior EV ownership also varies across quantiles. At the 80th 
percentile, respondents with prior EV ownership are 22.7 percentage points 
more likely to express strong purchase intent compared to those without prior 
ownership. This difference decreases to 20.8 and 16.2 percentage points at 
the median and 20th percentile, respectively. A similar pattern is 
observed in both the 2022 and 2023 surveys. Across surveys and quantiles, 
prior EV ownership is the strongest predictor for EV purchase intent.

Geographic and demographic factors show smaller but consistent relationships
with EV purchase intent across all surveys. Younger respondents (under 50), 
metropolitan residents, and those living in the West are more likely to 
express higher purchase intent across all considered quantiles, with the 
strongest covariate effects observed at the 80th percentile. 

Finally, the relationship between political party affiliation and EV purchase 
intent varies considerably across quantiles. Results from the 
2021 survey show that at the 80th percentile, Democrats are 14.3 percentage 
points more likely than Republicans to express strong purchase intent, with 
the difference narrowing to 9.6 percentage points  at the 
20th percentile. Similarly, at the 80th percentile Democrats are 8.3 
percentage points less likely than Republicans to express no interest in EVs, 
but the difference is 12.4 percentage points at the 20th percentile. These 
patterns remain consistent in more recent surveys. The results indicate that 
political identity is closely linked to EV purchase intent across the entire 
distribution: at higher quantiles, Democrats are significantly more likely to 
express strong purchase intent, and at lower quantiles, they are far less 
likely to reject EVs outright.

%---------------------------  Figure  ---------------------------------------
\begin{figure*}[!b]
	\centerline{
		\mbox{\includegraphics[width=7.00in, height=6.5in]{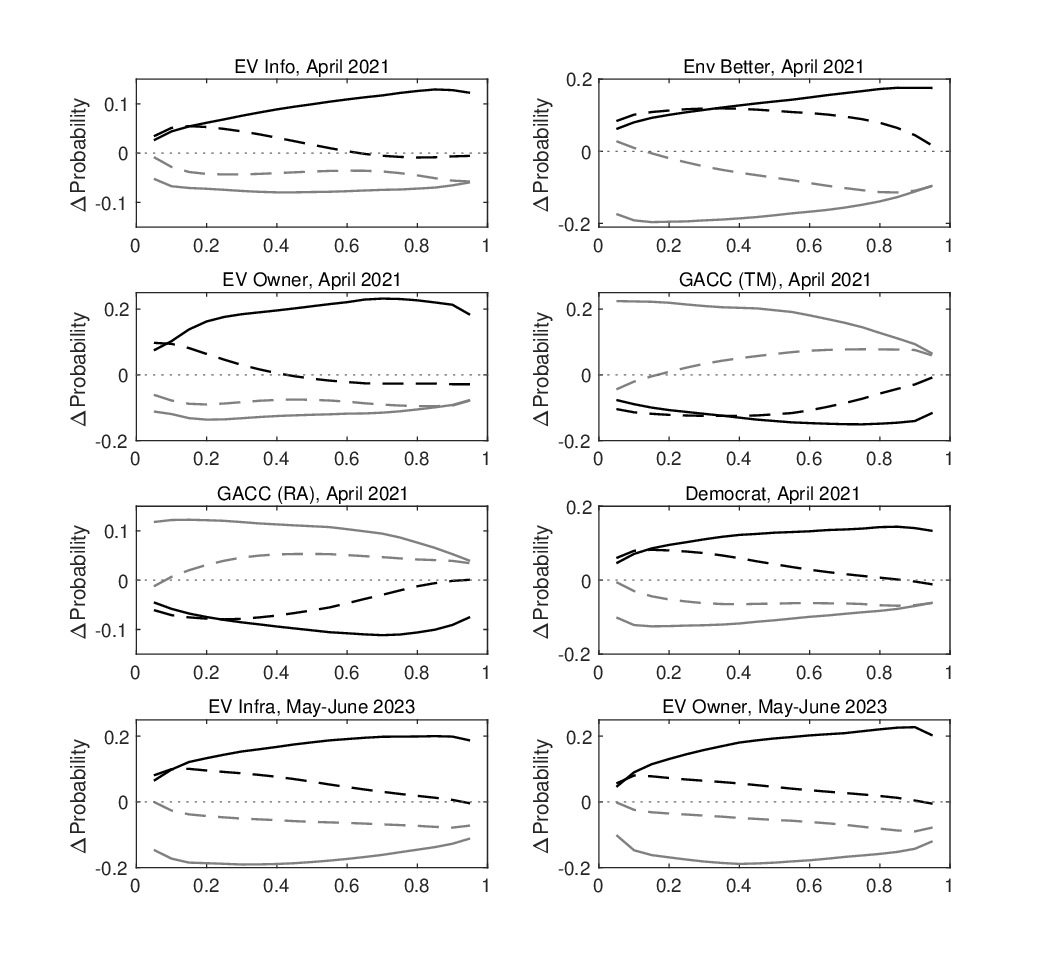}}
	}
	\vspace{-3pc}
	\caption{The figure presents the covariate effects for some selected 
		variables from the April 2021 and May-June 2023 surveys, spanning the 
		5th 
		to 95th quantiles with 5th quantile increments. Legend: $\Delta 
		\Pr(\textrm{Not at all 
			likely})$: \textcolor[rgb]{0.5, 0.5, 0.5}{\xdash}; 
		$\Delta \Pr(\textrm{Not too likely})$: \textcolor[rgb]{0.5, 0.5, 
			0.5}{\bxdash}\, \textcolor[rgb]{0.5, 0.5, 0.5}{\bxdash}; $\Delta 
		\Pr(\textrm{Somewhat likely})$: \bxdash\, \bxdash; $\Delta 
		\Pr(\textrm{Very likely})$: \xdash .
	}
	\label{fig:CEplot}
\end{figure*}
%-------------------------------------------------------------------------------

To provide a clearer understanding of how covariate effects vary across 
quantiles, Figure \ref{fig:CEplot} presents these for some 
selected variables from a series of ordinal quantile regression models, 
covering the 5th to 95th percentiles of EV purchase intent in 
5-percentile increments. Each panel corresponds to a different covariate; 
with the solid gray, dashed gray, dashed black, and solid black lines 
representing the change in probability of reporting ``not likely at all,'' 
``not too likely,'' ``somewhat likely,'' and ``very likely,'' respectively, 
for a given change in covariate across the quantiles. The figure clearly 
illustrates the substantial differences in 
covariate effects across the distribution.

For example, the solid black line indicates that respondents 
at higher quantiles who have read or heard extensively about EVs, perceive 
them as environmentally beneficial, have owned one, identify as 
Democrats, or are confident in the US ability to develop infrastructure are much more likely to report strong purchase interest compared to those at lower quantiles. Conversely, the solid 
gray lines show that those who believe government climate efforts are 
excessive or adequate are significantly more likely to report no interest at lower quantiles and markedly less likely to report strong 
interest at higher quantiles.

%------------------------------------------------------------------------------
\section{Discussion}\label{sec:Dicussion}
%------------------------------------------------------------------------------

Our findings point to several targeted strategies that may accelerate EV 
adoption by clarifying how beliefs and expectations shape purchase intent. 
First, exposure to EV-related information is associated with approximately a 
10 percentage point higher purchase intent. This pattern is notable given 
that fewer than 40 percent of respondents report high exposure, suggesting 
scope to expand the reach of EV-related information among prospective buyers. 
At the same time, smaller estimated effects at the lower end of the intent 
distribution indicate that increased awareness alone is unlikely to 
materially increase purchase intent among consumers with low baseline 
interest. Together, these results suggest that information provision may be 
most effective when targeted toward consumers already considering an EV. Such 
information could focus on practical considerations such as costs, 
incentives, and charging logistics. These results complement prior work that 
infers information exposure through peers or media \citep{Scherrer-2023, 
zhao2022determines} by relying on respondents' self-reported exposure and by 
highlighting heterogeneity across the intent distribution rather than average 
effects.

Second, belief that EVs have lower environmental impacts than gasoline 
vehicles is associated with higher purchase intent and lower reported 
disinterest. This result complements prior work linking pro-environmental 
attitudes to EV support \citep[e.g.,][]{hu2023policy, mamkhezri2025public} by 
showing that adoption decisions are shaped not only by general environmental 
orientation, but also by consumers' perceptions of EVs' environmental 
benefits relative to gasoline vehicles. It also extends evidence from 
smaller-sample studies examining the effects of perceived environmental 
benefits relative to gasoline vehicles to a larger, nationally representative 
US sample \citep{lashari2021consumers, Jia-etal-2025}. Survey responses 
indicate that a large majority of respondents believe EVs are environmentally 
preferable, but do not capture beliefs about the magnitude of those benefits. 
This measurement limitation prevents us from assessing how much additional 
adoption could be encouraged by shifting beliefs, and highlights a need for 
future research on how consumers perceive the size of environmental benefits 
and how those perceptions respond to information.

Third, confidence that sufficient charging infrastructure will be built is strongly associated with purchase intent, yet fewer than 20 percent of respondents report high confidence. Against the backdrop of prior work emphasizing physical access to charging infrastructure \citep{Carley-etal-2013, tiwari2020public}, our results point to expectations about the future charging network as a distinct channel shaping adoption decisions. These findings underscore the role of policy stability and credible long-term commitments in shaping expectations about the development of the charging network. Concerns about charger reliability and maintenance may also weaken confidence in future access even as new infrastructure is deployed. Such concerns were salient during the period covered by our survey data, and industry surveys reported that a substantial share of public-charging visits did not result in a successful charge due to charger unavailability or inoperability \cite{JDPower2023EVXCharging}. Policymakers and industry can help address this confidence gap through continued investment in charging infrastructure---such as through the National Electric Vehicle Infrastructure (NEVI) program---together with clear communication of deployment and maintenance plans.

Fourth, belief that the government is doing too much to address climate change is associated with lower EV purchase intent and is held by a non-trivial share of respondents. Our results complement prior findings linking EV interest to political affiliation \citep{mamkhezri2025public, sintov2020partisan} by indicating that evaluations of climate policy itself, in addition to partisan affiliation, shape EV adoption decisions. This pattern implies that framing EVs primarily as a climate policy instrument is unlikely to resonate with consumers who are skeptical of climate-focused policies. For these consumers, information about other EV attributes---such as operating costs, performance, or local air quality---that may be less salient ex ante can still influence adoption decisions. Moreover, information conveyed by non-governmental actors may carry greater credibility than government-led messaging.

Finally, prior EV ownership is strongly associated with stated purchase intent, consistent with learning or reduced uncertainty following initial adoption. This pattern suggests that policies lowering barriers for first-time buyers may influence adoption dynamics beyond the initial purchase. At the same time, systematic differences in stated interest across demographic groups may point to disparities in exposure, access, or information, highlighting the importance of inclusive policy design and outreach.

%------------------------------------------------------------------------------
\section{Conclusion}\label{sec:Conclusion}
%------------------------------------------------------------------------------

Electric vehicles (EVs) are widely viewed as a cornerstone of transportation 
decarbonization and climate change mitigation, yet 
federal support for EV adoption in the US has fluctuated sharply 
across administrations, reflecting broader partisan divides. In this context, 
consumer purchase sentiment plays a critical role in shaping both market 
outcomes and the political feasibility of sustained EV policy. Although 
public opinion is known to influence US policy making, existing research has 
not examined EV purchase intent using nationally representative data over 
multiple years, nor has it quantified how certain determinants operate across 
different levels of purchase interest. This study addresses these gaps by 
analyzing three nationally representative Pew Research Center surveys from 
2021$-$2023 using Bayesian ordinal probit and ordinal quantile regression 
models.

The empirical results indicate that EV purchase intent is positively 
associated with greater exposure to EV-related information, stronger beliefs 
in the environmental benefits of EVs, and confidence in the development of 
charging infrastructure. Political affiliation and demographic 
characteristics also play a significant role, with higher intent observed 
among younger, more educated, and higher-income individuals, metropolitan 
residents, Democrats, and those living in the western United States. In 
contrast, lower purchase intent is associated with skepticism toward 
government climate action, as well as with certain demographic 
characteristics, underscoring persistent heterogeneity in attitudes toward EV 
adoption across population subgroups.

The ordinal quantile analysis further reveals substantial heterogeneity 
in the determinants of EV purchase intent across the distribution. Greater EV 
knowledge, perceptions of environmental benefits, confidence in charging 
infrastructure, prior EV ownership, and Democratic Party affiliation increase 
the likelihood of strong purchase interest at higher quantiles while 
simultaneously reducing outright rejection at lower quantiles, whereas 
perceptions that the government is overreaching on climate policy are 
associated with lower interest throughout the distribution. These results 
provide a more nuanced perspective than mean-based ordinal models and yield 
clear policy and market insights: confidence in charging infrastructure 
emerges as one of the strongest predictors of adoption despite being reported 
by fewer than 20\% of respondents, underscoring the importance of 
infrastructure investment and transparent communication about availability 
and reliability. Similarly, broader exposure to EV-related information and 
messaging emphasizing environmental benefits---such as improved air quality---are 
associated with higher intent and reduced disinterest, while outreach that 
stresses cost, convenience, or energy security, particularly from 
non-government sources, may be more effective among climate-policy skeptics.

%------------------------------ Bibliography ---------------------------------
\clearpage \pagebreak
%\nocite{*}
\pdfbookmark[1]{References}{unnumbered}       % This makes References appear as a bookmark in pdf
%\section*{References}

\bibliographystyle{jasa}
\bibliography{EVBib}

\end{document}